\documentclass[%
superscriptaddress,
reprint,
nofootinbib,
 amsmath,amssymb,
aps,
prx,
]{revtex4-2}

\usepackage{graphicx}% Include figure files
\usepackage{dcolumn}% Align table columns on decimal point
\usepackage{bm}% bold math
\usepackage{dsfont}
\usepackage[
    colorlinks=true,
    linkcolor=blue,
    urlcolor=blue]{hyperref}% add hypertext capabilities

\DeclareUnicodeCharacter{2212}{\textendash}
\usepackage[T1]{fontenc}

\usepackage{xcolor}

\usepackage[noarrows]{ezedits}
\defineEdit{DK}{\color[rgb]{1.0,0.0,0.0}}{\color[rgb]{1.0,0.0,0.0}}
\defineEdit{SP}{\color[rgb]{0.0,0.0,1}}{\color[rgb]{0.0,0.0,1}}

\defineEdit{NLF}{\color[rgb]{1.0,0.53,0.0}}{\color[rgb]{1.0,0.53,0.0}}
\defineEdit{GL}{\color[rgb]{0.5,0.2,0.8}}{\color[rgb]{0.5,0.2,0.8}}
\defineEdit{DGM}{\color[rgb]{0.0,0.60,0.60}}{\color[rgb]{0.0,0.60,0.60}}
\defineEdit{MS}{\color[rgb]{0.05,.8,0.1}}{\color[rgb]{0.1,.8,0.1}}

\begin{document}
%\linenumbers

\title{Searching for dark matter with a 1000 km baseline interferometer}

\author{Daniel Gavilan-Martin}
\thanks{Both authors contributed equally.}
\affiliation{Johannes Gutenberg-Universit{\"a}t Mainz, 55128 Mainz, Germany}
\affiliation{Helmholtz Institute Mainz, 55099 Mainz, Germany}
\affiliation{GSI Helmholtzzentrum für Schwerionenforschung GmbH, 64291 Darmstadt, Germany}

\author{Grzegorz Łukasiewicz}
\thanks{Both authors contributed equally.}
\affiliation{Marian Smoluchowski Institute of Physics, Jagiellonian University in Krak\'ow, Łojasiewicza 11, 30-348, Krak\'ow, Poland}
\affiliation{Doctoral School of Exact and Natural Sciences, Jagiellonian University in Kraków, Łojasiewicza 11, 30-348, Krak\'ow, Poland}

\author{Mikhail Padniuk}
\affiliation{Marian Smoluchowski Institute of Physics, Jagiellonian University in Krak\'ow, Łojasiewicza 11, 30-348, Krak\'ow, Poland}

\author{Emmanuel Klinger}
% \affiliation{Johannes Gutenberg-Universit\"at Mainz, 55128 Mainz, Germany}
%  \affiliation{Helmholtz-Institut Mainz, GSI Helmholtzzentrum f{\"u}r Schwerionenforschung, 55128 Mainz, Germany}
 \affiliation{Institut FEMTO-ST -- UMR 6174 CNRS, SupMicroTech-ENSMM, Universit\'e de Franche-Comt\'e, 25030 Besan\c{c}on, France}

\author{Magdalena Smolis}
\affiliation{Marian Smoluchowski Institute of Physics, Jagiellonian University in Krak\'ow, Łojasiewicza 11, 30-348, Krak\'ow, Poland}

\author{Nataniel L. Figueroa}
\affiliation{Johannes Gutenberg-Universit{\"a}t Mainz, 55128 Mainz, Germany}
\affiliation{Helmholtz Institute Mainz, 55099 Mainz, Germany}

\author{Derek F. Jackson Kimball}
\affiliation{Department of Physics, California State University – East Bay, Hayward, CA 94542, USA}

\author{Alexander~O.~Sushkov}
\affiliation{Department of Physics, Boston University, Boston, MA 02215, USA}
\affiliation{Department of Electrical and Computer Engineering, Boston University, Boston, MA 02215, USA}
\affiliation{Photonics Center, Boston University, Boston, MA 02215, USA}
\affiliation{Department of Physics \& Astronomy, The Johns Hopkins University, Baltimore, MD  21218, USA}

\author{Szymon Pustelny}
\affiliation{Marian Smoluchowski Institute of Physics, Jagiellonian University in Krak\'ow, Łojasiewicza 11, 30-348, Krak\'ow, Poland}

\author{Dmitry Budker}
\affiliation{Johannes Gutenberg-Universit{\"a}t Mainz, 55128 Mainz, Germany}
\affiliation{Helmholtz Institute Mainz, 55099 Mainz, Germany}
\affiliation{GSI Helmholtzzentrum für Schwerionenforschung GmbH, 64291 Darmstadt, Germany}
\affiliation{Department of Physics, University of California, Berkeley, CA 94720-7300, United States of America}
 
\author{Arne Wickenbrock}
\affiliation{Johannes Gutenberg-Universit{\"a}t Mainz, 55128 Mainz, Germany}
\affiliation{Helmholtz Institute Mainz, 55099 Mainz, Germany}
\affiliation{GSI Helmholtzzentrum für Schwerionenforschung GmbH, 64291 Darmstadt, Germany}

\begin{abstract}
Axion-like particles (ALPs) arise from well-motivated extensions to the Standard Model and could account for dark matter. ALP dark matter would manifest as a field oscillating at an (as of yet) unknown frequency. The frequency depends linearly on the ALP mass and plausibly ranges from $10^{-22}$ to $10$\,eV/$c^2$. This motivates broadband search approaches. We report on a direct search for ALP dark matter with an interferometer composed of two atomic K-Rb-$^3$He comagnetometers, one situated in Mainz, Germany, and the other in Krak\'ow, Poland. We leverage the anticipated spatio-temporal coherence properties of the ALP field and probe all ALP-gradient-spin interactions covering a mass range of nine orders of magnitude. No significant evidence of an ALP signal is found. We thus place new upper limits on the ALP-neutron, ALP-proton and ALP-electron couplings reaching below $g_{aNN}<10^{-9}$\,GeV$^{-1}$, $g_{aPP}<10^{-7}$\,GeV$^{-1}$ and $g_{aee}<10^{-6}$\,GeV$^{-1}$, respectively. These limits improve upon previous laboratory constraints for neutron and proton couplings by up to three orders of magnitude.
\end{abstract}
%We use the ALP dark matter's spatiotemporal coherence properties assuming the standard halo model of dark matter in the Milky Way to improve the sensitivity and exclude spurious candidates.
\maketitle

\section{Introduction}
\begin{figure*}[htb]
    \centering
    \includegraphics[width=\textwidth]{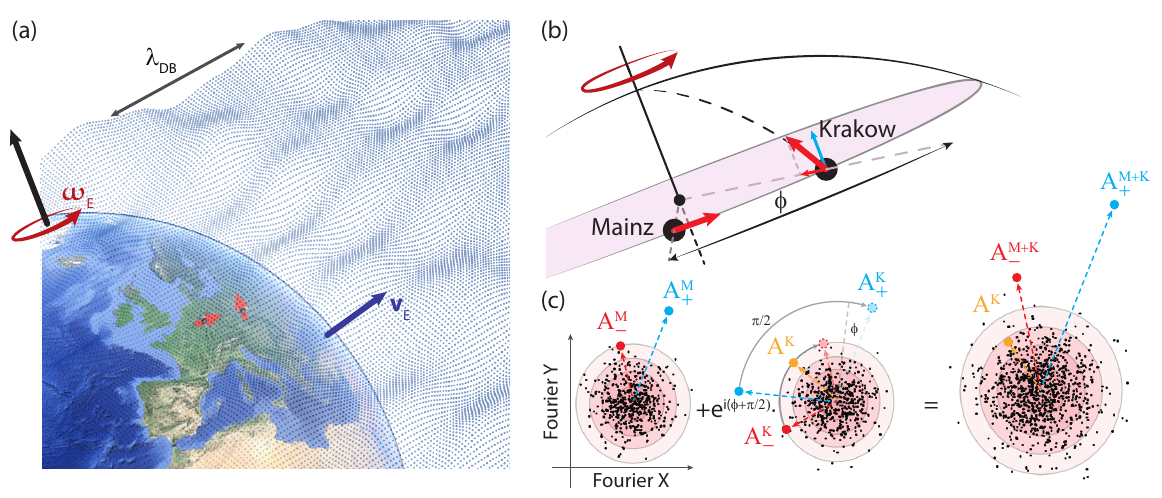}
        \caption{(a) Schematics of the comagnetometric interferometer. The two devices comprising the interferometer are indicated in their respective locations in Mainz and Krak\'ow. The small red arrows at the Earth surface point in the directions of the sensor sensitivity axes. In our model, Earth is moving in the galactic rest frame at velocity $v_{\text{E}}$ through the ALP DM field, characterized by its de Broglie wavelength $\lambda_{\text{DB}}$, that is more than a thousand times larger than the Earth radius. Besides its translational motion, Earth rotates around its axis giving rise to sidereal modulation of the signal at the frequency $\omega_{\text{E}}$.  (b) Orientation of the sensitive axes of the comagnetometers with respect to Earth's rotation axis. The sensitive axes of the comagnetometers can be decomposed into components along and perpendicular to the Earth's rotation axis. The former results in an ALP signal component (carrier) arising at the ALP Compton frequency $\omega_a$, while the latter results in (generally asymmetric) sidebands separated from the carrier by the sidereal frequency $\omega_\text{E}$. (c) Signal interferometry in the data analysis (illustration). The points in the three subfigures correspond to the complex Fourier amplitudes of all probed frequency bins of the Mainz (left), Krak\'ow (middle) datasets and their combination (right). We assume normal noise distributions. The circles indicate the standard deviation. The points marked with red, blue, and orange represent injected ALP signatures observable in both comagnetometers and how they appear in the combined signal. Due to the directional sensitivity of the comagnetometers, the injected ALP signal manifests as a carrier of amplitude $A^\text{K}$ only in the Krak\'ow data, and sideband signatures of different amplitudes and phases in both Krak\'ow ($A_\pm^\text{K}$) and Mainz ($A_\pm^\text{M}$) data. The phase difference between the signals arises due to the different orientation of the sensitive axes ($\pi/2$), as well as the different locations ($\phi$) of the sensors. Appropriate phasing allows to coherently add the ALP signals, while the noise adds incoherently.}
    \label{fig:schematic}
\end{figure*}
%DM - What is it, why is it important?
Abundant astrophysical and cosmological observations at different scales~\cite{Bertone2005Jan,Feng2010Sep, Gorenstein2014Jun} suggest that about $85\%$ of the matter in the Universe does not noticeably interact beyond  gravitational interactions and is thus known as ``dark'' matter (DM). Since the DM composition remains unknown, it serves as a provocative indication of physics beyond the Standard Model (SM) and drives the search for hypothetical DM particles.

Ultralight ($\, \lesssim 10$\,eV/$c^2$) pseudoscalar bosons
%\SP{Can we replace pseudoscalar with bososn or pseudoscalar bosons?} 
known as axion-like particles (ALPs) are particularly well-motivated DM candidates~\cite{Antypas2022Mar, abbott_cosmological_1983, preskill_cosmology_1983, dine_not-so-harmless_1983}. ALPs could account for the correct abundance of relic DM (via, for example, a misalignment mechanism \cite{co_axion_2020}) and a subset of them could resolve the strong-CP problem~\cite{Marsh2016Jul}. Additionally, ALPs can interact with SM particles through couplings to photons, gluons, and fermions~\cite{Graham2013Aug}, offering direct ways of probing their existence~\cite{budker_proposal_2014,jackson_kimball_search_2023,graham_experimental_2015}.

In this article, we present the results of an interferometric ALP search (see also Ref.\,\cite{Safdi_2021_Interferometer,crescini_fermionic_2023}). %\SP{This sounds as our search is somehow associated with the other search. What if we take reference and put at the end "(see also Ref.\,\cite{}"? This does not suggest they are related}
The search explores a wide mass range (nine orders of magnitude) and investigates ALP couplings to three distinct spin types (i.e., those associated with proton, neutron, and electron).  The interferometer is composed of two K-Rb-$^3$He atomic comagnetometers, one located at the Jagiellonian University in Krak\'ow, Poland \cite{Padniuk2024PhD} and the other, separated by 860\,km, at the Johannes Gutenberg University in Mainz, Germany, see Fig.\,\ref{fig:schematic}(a). The interference occurs through the phase-sensitive combination of the amplitude data, in a manner similar to that used in radio telescope networks (e.g., the event horizon telescope \cite{collaboration_first_2019}). The main difference between our and other schemes is that we are sensitive to ALP DM gradients rather than electromagnetic waves and that the corresponding improvement in angular resolution does not play a role in searching for DM, which is assumed to have a homogeneous local density. It might, however, be important in searches when DM models feature more heterogeneous mass distribution including streams and halos \cite{Budker_2023_generic,OHare_Axion_2024,kryemadhi_gravitational_2023, khlopov_nonlinear_1999}, or if there are distinguishable emitters of bosonic fields in the Universe, e.g., black holes emitting ALPs in a process of superradiance \cite{Baryakhtar_Black_2021} or during binary merger \cite{dailey_quantum_2020}.

We optimize both comagnetometers  to operate in the self-compensating regime \cite{kornack_dynamics_2002,wei_ultrasensitive_2023,klingerPRA2023}. To a first order, the devices are insensitive to low-frequency magnetic-field variations, but retain sensitivity to non-magnetic spin interactions~\cite{ padniuk_response_2022,padniuk_universal_2024}. This makes them excellent tools for probing the interaction of the galactic ALP DM field $a(\bm{r},t)$ with neutron spins $\bm{\sigma}_N$, proton spins $\bm{\sigma}_P$ and electron spins $\bm{\sigma}_e$, which are described by the Hamiltonians
\begin{equation}
    \begin{split}
    \mathcal{H}_N &= g_{aNN} \bm{\nabla}a \cdot \bm{\sigma}_N\,,\\
    \mathcal{H}_P &= g_{aPP} \bm{\nabla}a \cdot \bm{\sigma}_P\,,\\
    \mathcal{H}_e &= g_{aee} \bm{\nabla}a \cdot \bm{\sigma}_e\,,
    \end{split}
\end{equation}
%\DGM{Check missing $\gamma$ and $\mu$}
where $g_{aNN}$, $g_{aPP}$, and $g_{aee}$ are unknown coupling constants to neutrons, protons and electrons. Self-compensating comagnetometers have already established the most stringent limits in certain mass ranges, even surpassing astrophysical constraints \cite{lee_laboratory_2023,wei_dark_2023,xu_constraining_2024}. In this work, we search in experimentally-unconstrained ALP parameter space in the ultralight mass range below $10^{-13}$\,eV/$c^2$.

If the estimated local DM density ($\approx0.4$\,GeV/cm$^3$)  \cite{Workman_2022_PDG} is mostly due to an ALP of mass $m_a$, the occupation number of the ALP field would be large and hence it can be approximated as a classical field oscillating near the ALP Compton frequency \cite{chadha-day_axion_2022}. 
In this model, the characteristics of the oscillating ALP field, such as the amplitudes and the phases of the ALP gradient components, fluctuate in time according to the properties of virialized dark matter \cite{foster_revealing_2018, centers_stochastic_2021, lisanti_stochastic_2021, gramolin_spectral_2022, flambaum_fluctuations_2023}. The characteristic time scale of the fluctuations results from the ALP DM velocity spread, which leads to a coherence time of around $10^6$ oscillations of the field (for example, about 15 years for $m_a = 10^{-17}$\,eV/$c^2$). 

In our work, we focus primarily on the regime where the coherence time of the ALP collective oscillation is larger than the total measurement time. In this regime, the ALP field can be treated as having a constant amplitude and direction and hence its signatures at the two sensor locations are highly correlated. For reference, an ALP particle of mass $10^{-17}$\,eV/$c^2$, assuming a relative velocity equal to that of Earth in the galactic rest frame, has a de Broglie wavelength of $\sim10^3$ astronomical units. Thus, by properly combining the signals from both stations, the ALP signals will be added constructively, while the effects of local noise fluctuations sum incoherently and are suppressed, see Fig.\,\ref{fig:schematic}(c). 

An additional analysis of the results allows us to extend the search to ALP masses below $\sim 5 \times 10^{-20}$\,eV/$c^2$. This is possible by looking for spectral signatures in the data that would arise due to the rotation of Earth.  The properties of the field gradient in this ultralow frequency regime are discussed and the signal properties are explicitly reviewed below.

This article is structured as follows. In Sec.\,\ref{sec:DM_model}, we review the ALP DM signal model, explicitly showing the expected signatures in the frequency domain for a daily modulated sensor, and discuss how the sensitivity can be improved by analyzing interfering signals from multiple sensors. In Sec.\,\ref{sec:SearchResults}, we show the analysis framework and discuss how data is combined to maximize a signal-to-noise ratio. As no ALP candidates are found, we set limits on the ALP DM pseudoscalar spin interactions in Sec.\,\ref{sec:limits}. The experimental setup and some technical details of the search are presented in Sec.\,\ref{sec:methods}.

% 581 words

\section{Signal model}\label{sec:DM_model}

To describe the expected-signal model in the two comagnetometer locations, we utilize the framework described in Refs.\,\cite{lee_laboratory_2023, gramolin_spectral_2022, lisanti_stochastic_2021}. Due to its high occupation number, the ALP-field can be approximated as a sum of $N$ independent oscillations. We write the components of the ALP field gradient at position $\boldsymbol{r}$~and time $t$ in a  Cartesian frame of reference $i = x,y,z$, where the Solar System is at rest
\begin{equation}\label{eq:gradientsum}
    \begin{split}
\boldsymbol{\nabla}_i a(\boldsymbol{r},t) & =  \frac{\hbar\sqrt{2\rho_{\textrm{DM}}}}{m_ac\sqrt{N}}\sum_{n=1}^N \boldsymbol{\nabla}_i \sin(\omega_nt-\boldsymbol{k}_n\cdot \boldsymbol{r} + \phi_n') \\
  & = \frac{\sqrt{2\rho_{\textrm{DM}}}}{c\sqrt{N}}\sum_{n=1}^N (\boldsymbol{v}_{n})_i \cos(\omega_n t - \boldsymbol{k}_n\cdot \boldsymbol{r} + \phi_n)\,.
    \end{split}
\end{equation} 
The negative sign in the second line is absorbed by the random phase $\phi_n=\phi_n'+\pi$, $c$ is the speed of light, $m_a$ is the ALP mass, $\rho_\text{DM}\approx 0.4$\,GeV/cm$^3$ \cite{Workman_2022_PDG} is the local DM density, $\boldsymbol{k}_n = m_a\boldsymbol{v}_n/\hbar$ is the ALP-field wave vector, $\omega_n$ is the angular frequency of each oscillating mode $n$,~and $(\boldsymbol{v}_n)_i$ is the $i$-th component of the relative velocity $\boldsymbol{v}_n$ between the ALP mode and the sensor. Assuming that ALPs are virialized in the Milky Way, the velocities $\boldsymbol{v}_n$ follow a 3D normal distribution centered at zero in the galactic frame. On Earth, the observed velocity distribution will be offset by Earth's velocity in the galactic frame $(\boldsymbol{v_{\text{E}}})_i$. We neglect effects due to the movement of Earth within the Solar System as it is negligible compared to the velocity of the Solar System in the galaxy \cite{gramolin_spectral_2022}. The velocity components $(\boldsymbol{v}_n)_i$ are then distributed according to
\begin{equation}
\label{eq:velocity_distribution}
    f(v_i) =  \frac{1}{v_0\sqrt{\pi}}\exp\left\lbrace-\left[\frac{v_i-(\boldsymbol{v_{\text{E}}})_i}{v_0} \right]^2\right\rbrace,
\end{equation}
where $v_0 \approx 220$\,km/s is the virial velocity in the Milky Way, determining the variance of the velocity distribution to be $v_0^2/2$.

Each mode of the ALP field experiences a kinetic energy correction, leading to a small frequency shift \hbox{$\omega_n \approx \omega_a(1+\boldsymbol{v}_n^2/2c^2)$}. Thus, most of the ALP spectrum is concentrated within a spectral width given by \cite{gramolin_spectral_2022}
\begin{equation}
    \Delta \omega \approx \omega_a \frac{v_0^2}{c^2} \approx \omega_a \times 10^{-6}.
\end{equation} 

 The collective mode can be described by a nearly monochromatic oscillation with independent amplitude and phase for each orthogonal spatial direction. The amplitudes and phases are random and evolve smoothly on time scales set by the coherence time $\tau\approx 2\pi/\Delta \omega$. The underlying probability distributions of the amplitudes and phases can be derived from Eqs.\,\eqref{eq:gradientsum} and \eqref{eq:velocity_distribution} \cite{gramolin_spectral_2022}. This results in six independent parameters: three amplitudes ($\alpha_x$, $\alpha_y$, and $\alpha_z$), following Rayleigh distributions, and three phases ($\phi_x$, $\phi_y$, and $\phi_z$), following  uniform distributions over $[0,2\pi)$ \cite{lisanti_stochastic_2021}. Because the ALP de Broglie wavelength in the analyzed mass range is much larger than the sensor separation, $\lambda = 2\pi/k_n \gg d = 860$\,km, we neglect the spatial dependence, i.e., $\boldsymbol{k}_n\cdot \boldsymbol{r}\approx 0$,  in Eq.\;\eqref{eq:gradientsum} for both sensors. 
The ALP gradient can then be written as
% from SP: comment that the summation results in Rayleigh distribution
\begin{align}\label{eq:DMfield3D}
\begin{split}
    \boldsymbol{\nabla} a (t)  = \frac{\sqrt{2\rho_{\textrm{DM}}}}{c\sqrt{N}} \sum_{n=1}^N \boldsymbol{v}_n \cos&(\omega_a t + \phi_n) \\ 
     = \boldsymbol{\hat{x}}\alpha_x \cos(\omega_a t + \phi_x) &+ \boldsymbol{\hat{y}}\alpha_y \cos(\omega_a t + \phi_y) \\
     &+ \boldsymbol{\hat{z}}\alpha_z \cos(\omega_a t + \phi_z) \,,
    \end{split}
     \end{align}
where the sum can be evaluated with the central limit theorem, and results in the amplitude terms $\alpha_i$.
The probability distributions of the amplitudes $\alpha_i$ are given by \cite{gramolin_spectral_2022}
\begin{equation}  
\label{eq:alphas}
\alpha_i \sim \frac{\sqrt{2\rho_{\textrm{DM}}}}{c} \sqrt{\frac{v_0^2/2+(\boldsymbol{v_{\text{E}}})_i^2}{2}} \alpha\,. 
\end{equation}
where $\alpha$ is a Rayleigh distributed random number with scale parameter equal to 1.

The sensitive axis $\boldsymbol{\hat{m}}$ of a single sensor located at Earth's surface rotates with the sidereal frequency of Earth, $\omega_{\text{E}}$. The coordinate system is chosen such that the $\boldsymbol{\hat{z}}$ component is parallel to the Earth rotation axis. The coordinate system can be assumed static in the galactic rest frame over the timespan of the experiment. This results in a~fixed $m_z$ component and daily modulated $m_x$ and $m_y$ components: 
\begin{equation}
\begin{split}
    \boldsymbol{\hat{m}}(t) & = \boldsymbol{\hat{x}} \sin\theta \sin(\omega_{\text{E}} t + \phi_{\text{E}})\\ 
    &+ \boldsymbol{\hat{y}} \sin\theta \cos(\omega_{\text{E}} t + \phi_{\text{E}}) \\
    &+ \boldsymbol{\hat{z}} \cos\theta\,,
\end{split}
\end{equation}
where $\theta = \angle(\boldsymbol{\hat{z}}, \boldsymbol{\hat{m}})$ is the polar angle and $\phi_{\text{E}}= \angle(\boldsymbol{\hat{x}}, \boldsymbol{\hat{m}})$ is the azimuthal angle in a spherical Earth coordinate system. The experimental signal is proportional to the projection of the gradient of the ALP field on the sensitive axis of the sensor,

\begin{widetext}
\begin{equation}
\label{eq:modulated_signal}
\begin{split}
\boldsymbol{\nabla} a(t) \cdot \boldsymbol{\hat{m}}(t)
 =~& \frac{\alpha_x\sin\theta}{2}\Big\lbrace \sin\big[(\omega_a+\omega_{\text{E}})t+\phi_x+\phi_{\text{E}}\big] - \sin\big[(\omega_a-\omega_{\text{E}})t+\phi_x-\phi_{\text{E}}\big] \Big\rbrace \\
 + & \frac{\alpha_y\sin\theta}{2}\Big\lbrace \cos\big[(\omega_a+\omega_{\text{E}})t+\phi_y+\phi_{\text{E}}\big] + \cos\big[(\omega_a-\omega_{\text{E}})t+\phi_y-\phi_{\text{E}}\big] \Big\rbrace \\
+ & \alpha_z\cos\theta\cos(\omega_at+ \phi_z)\,.
\end{split}
\end{equation}
\end{widetext}

The pseudomagnetic field in the experiment is given by $B_a=\frac{g_\text{eff}}{\mu_n}\boldsymbol{\nabla} a \cdot \boldsymbol{\hat{m}}$,
%\DGM{Check missing factors},
with $\mu_n/2\pi = -32.4$\,MHz/T the gyromagnetic ratio of $^3$He, and $g_\text{eff}= g_{aNN} \xi_N\, (g_{aPP} \xi_P)$ the effective nucleon coupling taking into account the neutron (proton) content $\xi_N\, (\xi_P)$ \cite{kimball_nuclear_2015}. In the case of an electron interaction, the effective coupling is given by the ratio of the gyromagnetic ratios of electron and neutron $g_\text{eff}= g_{aee} \gamma_n/\gamma_e$ (see Sec.\,\ref{sec:methods-electron}). 

To obtain the frequency-domain signature of Eq.\,\eqref{eq:modulated_signal}, we describe the three frequency components of the amplitude-modulated signal as a carrier, oscillating at the ALP Compton frequency $\omega_a$ and two sidebands, separated from the carrier by Earth's sidereal frequency $\pm\omega_{\text{E}}$. 
The magnitudes of the three components (in magnetic-field units) are given by
\begin{equation}\label{eq:amplitudes}
    \begin{split}
    &|A| =\frac{g_\text{eff}}{\mu_n} \alpha_z \cos\theta  \, , \\
    &|A_-| =\frac{g_\text{eff}}{\mu_n}\frac{\sin\theta}{2}\sqrt{ \alpha_x^2 + \alpha_y^2 - 2\alpha_x\alpha_y\sin(\phi_x-\phi_y) }\, , \\
    &|A_+|  = \frac{g_\text{eff}}{\mu_n} \frac{\sin\theta}{2}\sqrt{ \alpha_x^2 + \alpha_y^2 + 2\alpha_x\alpha_y\sin(\phi_x-\phi_y) } \, .
    \end{split}
\end{equation}
Because there are three complex amplitudes $A$ and $A_\pm$ that depend differently on six random variables $\alpha_i$ and $\phi_i$,
%they are independent of one another 
%Due to the fact that $\alpha_i$ and $\phi_i$ are independent random variables, 
the amplitudes are independent of one another.
It is worth noting that even though the signal in the different frequency components are a result of an amplitude modulation, the sideband amplitudes are generally asymmetric. However, for measurements that average a large number of coherence times, the sidebands magnitude converge to the same value.

Note that the total power of the gradient of the ALP DM signal within a single coherence patch distributed among three frequencies $\omega_a$, $\omega_+$, and $\omega_-$ is
\begin{equation}
\label{eq:ALP-power}
        \frac{|A|^2}{\cos^2\theta}+\frac{2(|A_+|^2+|A_-|^2)}{\sin^2\theta}= 
    g_\text{eff}^2(\alpha_z^2+\alpha_x^2+\alpha_y^2)=       g_\text{eff}^2 |\boldsymbol{\alpha}|^2\,,
\end{equation}
where $\boldsymbol{\alpha}=(\alpha_x,\alpha_y,\alpha_z)$ gives the 3D amplitude of the ALP DM field.

In our experimental search for ALP DM, it is essential to consider the role of noise in the comagnetometer data.
Correlated measurements of ALP DM with two sensors instead of a single sensor lead to a substantial improvement of sensitivity. 
In fact, the improvement is greater than a factor of $\sqrt{2}$ expected from two measurements of a quantity with uncorrelated noise \cite{derevianko_detecting_2018}. 
The reason lies in the respective orientation of the sensitive axes of the sensors. In our configuration, the combination of both sensors allows access to all spatial components of the ALP gradient signal. 

The individual ALP gradient components are independent random values described by Rayleigh distributions. 
 Probing all of them simultaneously using the ALP DM interferometer configuration enables us to increase the combined signal amplitude
 and extend the range of the search into lower values of $g_{\textrm{eff}}$.

The spatial configuration of the two-station ALP DM interferometer is as follows: the sensitive axis of the Mainz station is horizontal in the laboratory pointing East  in the Earth rotation plane ($\theta_M=90\,^\circ$), and therefore, is  exclusively sensitive to the sideband signal $A_\pm$. In contrast, the sensitive axis of the Krak\'ow station is horizontal in the laboratory pointing North ($\theta_K=50\,^\circ$). The Krak\'ow comagnetometer is therefore sensitive to all ALP spectral signatures (carrier and sidebands).
% 436 words excluding equations 

\section{Search strategy and results} \label{sec:SearchResults}
In our experiment, we searched for ALP DM with Compton frequencies up to 11.6\,Hz. In this frequency range, the linewidth of the ALP signal is below the frequency resolution of the analysis method.
Additionally, the sensitive axes of the comagnetometers are modulated at the Earth sidereal frequency, $\omega_{\text{E}}/2\pi \approx 11.6\,\mu$Hz, resulting in resolvable sidebands, see Eqs.\,\eqref{eq:amplitudes}. The expected amplitude ratio between the central peak (carrier) and sidebands is given by the components of the sensitive axes parallel (carrier) and perpendicular (sidebands) to the rotation axis of Earth. 

The data analyzed in this work correspond to a total of 40 measurement days in Mainz and 28 days in Krak\'ow, collected  over a 92-day period between January~6 and April 7, 2024. Data collection was performed as consistently as possible, but several technical interruptions occurred during the measurement campaign.
From that data, 25-hour continuous data segments were used for the analysis, ensuring sidereal frequency resolution ($\Delta \omega \lesssim \omega_{\text{E}}$). 

In general, when searching for unknown dynamic-observables, it is crucial to precisely know the sensor frequency response. However, in the case of ALP DM, the frequency response cannot be directly measured. To address this issue, the frequency responses of both comagnetometers to exotic interactions were inferred using the method reported in Ref.\,\cite{padniuk_universal_2024}. The method involves (1) measuring the response to a magnetic step perturbation, (2) fitting the Discrete Fourier Transform (DFT) data with a model that describes the coupled spin dynamics, and (3) inferring the frequency response of  neutrons, protons, and electrons to postulated non-magnetic (exotic) interactions. In our analysis, the frequency responses was determined for each 25\,h data segment. 
%were determined using  of the time series data and were subsequently used for device calibration over the period.
The calibration-fit results for Mainz and Krak\'ow are summarized and discussed in Sec.\,\ref{sec:methods-setup}.

To coherently combine ALP DM signals from different DFT subsets, a frequency-dependent shift is applied and the subsets are phase aligned.
%(see Sec.\,\ref{sec:methods})
Then, all DFT subsets of the respective station are averaged, resulting in a mean value and its standard deviation. We confirmed that the time-shifted DFT averaging procedure had no effect on the Fourier coefficients of injected oscillations. 

To search for ALP DM signals [Eq.\,\eqref{eq:ALP-power}], we construct an estimator of signal amplitude $S(\omega)$. To achieve the optimum signal-to-noise ratio, we extract all potential ALP DM signatures from the data by combining the Mainz and Krak\'ow DFT datasets. We then combine the power of carrier and sidebands. Specifically, the complex Fourier components of the sidebands ($ A_{\pm}$) in Mainz and Krak\'ow are added with weights taking into account the direction of the sensitive axes and the noise level (see Sec.\,\ref{sec:methods-weights}). The interfered sideband signal then reads
\begin{equation}
\label{eq:sideband_interference}
    A^{K+M}_{\pm}=\frac{ 
  a^M_{\pm}   A^{M}_{\pm}  + a^K_\pm A^{K}_{\pm} e^{-i(\phi+\pi/2)}  }
  {  a^K_{\pm} + a^M_{\pm} }\,,
\end{equation}
where $a^i_\pm$ are the weights with $\pm$ designating the higher ($+$) and lower ($-$) frequency sideband and superscript \hbox{$i=M,K$} indicates the Mainz and Krak\'ow stations, respectively. The angle between the positions of the Mainz and Krak\'ow comagnetometers in the rotation plane of the Earth is $\phi$, see Fig.\,\ref{fig:schematic}(b). Accounting for the angle $\phi$ is required to ensure constructive interference of an oscillating ALP DM signal recorded in the two comagnetometers. The constructive interference of the ALP DM signal in the combination of the sidebands is illustrated in Fig.\,\ref{fig:schematic}(c).
% such that phase difference between both stations is $phi_E^K - \phi_E^M = \phi+\pi/2 $

Next, we combine the carrier in Krak\'ow $A^K$ with the interfered sideband signal $A^{K+M}_{\pm}$. The signal estimator $S(\omega)$ for ALP DM incorporates the appropriate weights for averaging signal power; $b^K$ for the carrier in Krak\'ow and $b^{K+M}_{\pm}$ for the upper and lower interfered sideband, see Sec.\,\ref{sec:methods-weights}.
The Mainz carrier signal does not contribute to the signal estimator $S(\omega)$, as its weight $b^M \approx 0$, because $\cos(\theta_M)\approx 0$, meaning that there is no contribution to $A^{M}$ from ALP DM.
The estimator for the ALP DM signal $S(\omega)$ is defined in the following way
\begin{widetext}    
\begin{equation}\label{eq:SigEstimator}
  S (\omega) = \sqrt{\frac{b^K \left|A^K\right|^2 +  b^{K+M}_{-}\left|A^{K+M}_-\right|^2  + b^{K+M}_{+}\left|A^{K+M}_+\right|^2  }{b^K+ b^{K+M}_{-}+b^{K+M}_{+}}}\,.
\end{equation}
\end{widetext}
Note that the standard deviation of the DFTs is propagated accordingly, resulting in an uncertainty in the signal estimator $\Delta S(\omega)$ for each frequency.

\begin{figure}[htb]
    \centering

        \includegraphics[width=0.5\textwidth]{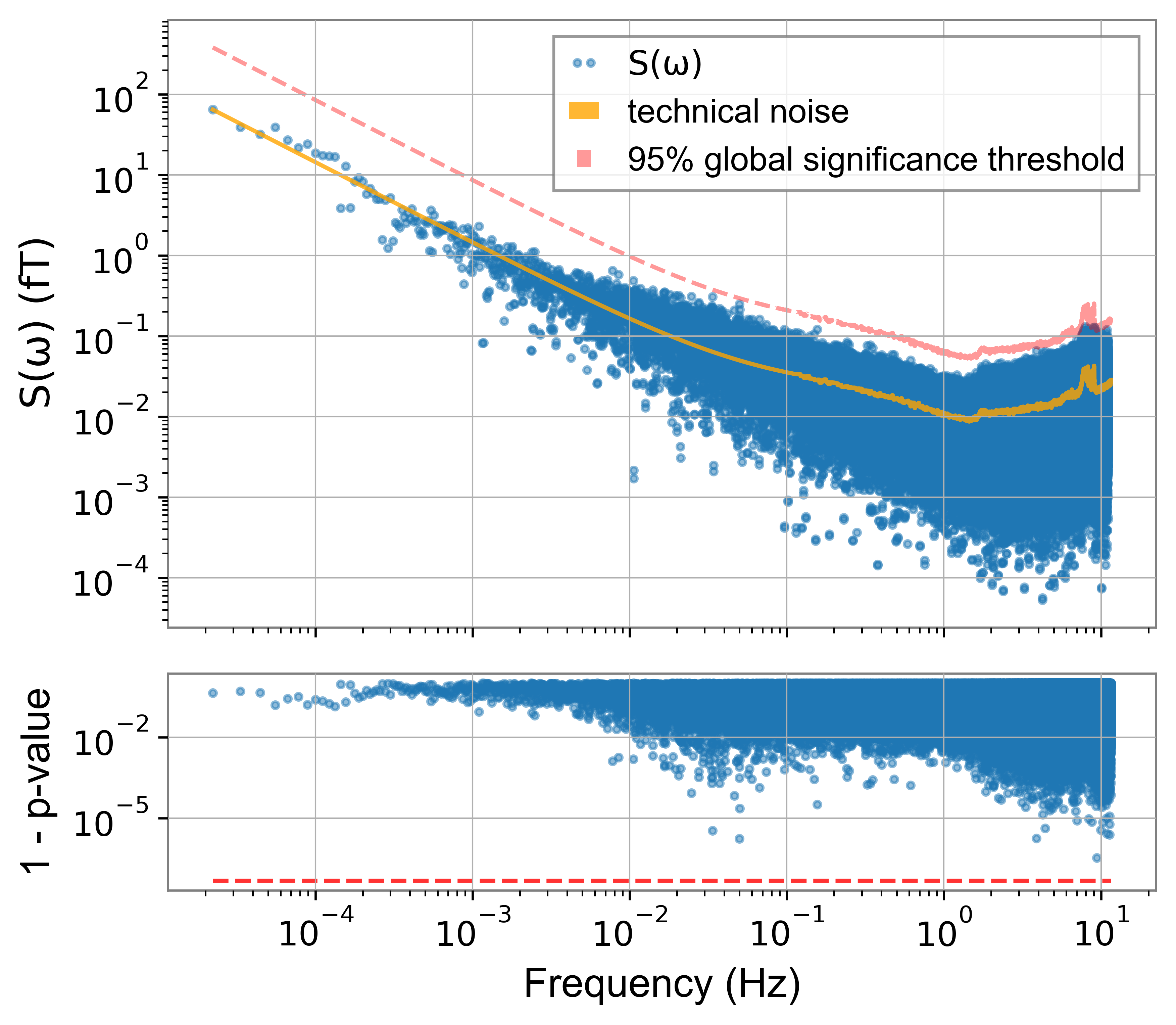}
        
        \caption{Signal estimator $S(\omega)$, obtained by interfering the Mainz and Krak\'ow comagnetometer data, for frequencies above $\omega_{\text{E}}$. The results are shown as a function of frequency in the upper figure. The data shows a $1/f$ scaling behaviour up to $10^{-1}$\,Hz consistent with technical noise of the apparatus. The peak sensitivity of the estimator reaches $10^{-17}$\,T. No ALP candidate is found beyond the global 95\% significance threshold in p-value, as shown in the lower plot.}
    \label{fig:detectionthreshold}
\end{figure}

The value of $S(\omega)$, given in magnetic-field units and determined from the acquired data, is represented  by the blue points in Fig.\,\ref{fig:detectionthreshold}. An ALP DM signal in the $S(\omega)$ data would correspond to a single data point at frequency $\omega_a$. Since this is one point out of the whole dataset consisting of a million points, one per frequency, the distribution of $S(\omega)$ over all frequencies characterizes the technical-noise of the interferometer.
%we assume that the distribution of$S(\omega)$ evaluated for all sampled frequencies characterizes the technical noise of our interferometer.

In order to estimate the expected value of the technical noise amplitude at each frequency, fit($\omega$), we use the measured values of $S(\omega)$ at surrounding points. For frequencies below 0.1\,Hz, the mean is inferred from a global fit assuming a $1/f$ noise model. Above 0.1\,Hz the mean is based on the moving average of 500 consecutive points centered around (but excluding) the frequency of interest. Figure~\ref{fig:detectionthreshold} shows the mean of the technical noise as an orange line determined as discussed above, as well as a light red line indicating the 95\% global significance threshold for each frequency.

To determine whether any of the estimator data points are significantly larger than expected from the measured technical noise, a detection threshold is established. The set of the signal-estimator values $S(\omega)$, normalized with the expected noise at each frequency, is found to match a non-central $\chi^2$ distribution with six degrees of freedom (see Sec.\,\ref{sec:methods:noise}). 
The fitted non-central $\chi^2$ distribution is used to set the detection threshold to guarantee that a candidate signal has only a 5\% global chance of arising due to technical noise which accounts for the look-elsewhere effect. The measured values of $S(\omega)$ are consistent
with noise and therefore shows no evidence of an ALP
DM candidate.

 \begin{figure}[htb]
    \centering

    \includegraphics[width=0.5\textwidth]{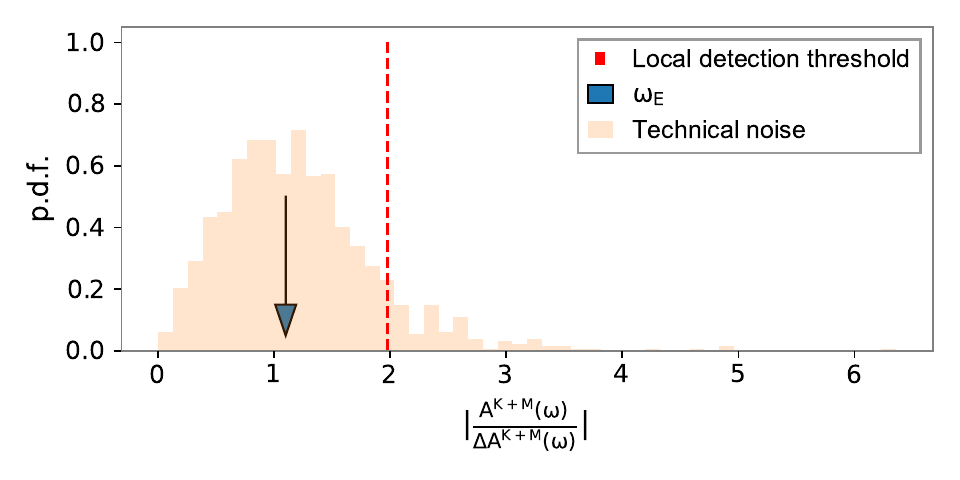}
        
        \caption{Histogram of the measured-noise distribution and the combined amplitude value at $\omega_{\text{E}}$ as a function of the measured Fourier amplitude. For an ALP field oscillating below $\omega_{\text{E}}$, the sidebands are not resolved and an ALP signature is present at $\omega_{\text{E}}$. However, $A^{M+K}(\omega_{\text{E}})$ is compatible with the expected noise distribution and thus no ALP candidates with frequencies $\omega_a<\omega_{\text{E}}$ are reported.  The blue arrow indicates the frequency bin at $\omega_{\text{E}}$, the dashed red line indicates the 95\% confidence detection threshold  and the orange distribution is the measured noise for the the frequency bins between $\omega_{\text{E}}$ and 0.01\,Hz. The amplitude is normalized by the standard deviation $\Delta A^{M+K}(\omega)$ to whiten the noise.
        }
    \label{fig:we}
\end{figure}

For ALP DM with frequencies below $\omega_{\text{E}}$, a different search approach is used. An ALP DM contribution at $\omega_{\text{E}}$ would be observed for the Compton frequencies $\omega_a < \omega_{\text{E}}$ even if $\omega_a$ cannot be directly resolved in our data. The width of each frequency bin is approximately equal to $\omega_{\text{E}}$. In this situation, the sidebands are at $\omega_{\text{E}} \pm \omega_a$. Since $\omega_{\text{E}}>\omega_a$, they appear in the same frequency bin at $\omega_{\text{E}}$. By examining the frequency bin at $\omega_{\text{E}}$, the ALP search can be extended to the limit where the ALP oscillation is much slower than the rotation of Earth ($\omega_a \ll \omega_{\text{E}}$). Due to the particular characteristics of this bin, we do not include it in the general ALP DM search described in the previous paragraph. Instead, we add Mainz and Krak\'ow amplitudes with their respective weights and obtain $A^{K+M}(\omega)$ following Eq.\,\eqref{eq:sideband_interference}. We use the estimator $A^{K+M}(\omega)/\Delta A^{K+M}(\omega)$ normalized by the standard deviation $\Delta A^{M+K}(w)$ as the expected noise model. For the frequency bins below 0.1 Hz the estimator histogram follows a Rayleigh distribution which is used to calculate detection threshold (95\% local significance threshold) for the measured value $A^{K+M}(\omega_{\text{E}})$. The results are shown in Fig.\,\ref{fig:we}. The measured amplitude at $\omega_{\text{E}}$ is $S(\omega_\text{E})=4.7 \times 10^{-14}$\,T and the standard deviation is $\Delta S(\omega_\text{E})=4.2 \times 10^{-14}$\,T. It is consistent with noise and therefore shows no evidence of an ALP DM candidate.

% 770 words 

\section{Setting limits} \label{sec:limits}
Since our measurements did not reveal any ALP DM candidates in the frequency range from $10^{-8}$\,Hz to 11.6\,Hz, we proceed to set limits on the ALP DM pseudoscalar couplings. Specifically, we set exclusions on the ALP-proton coupling $g_{aPP}$, ALP-neutron coupling $g_{aNN}$, and ALP-electron coupling $g_{aee}$. For simplicity, when constraining one of the three couplings, we do not consider interactions due to the other couplings. We describe the exclusion strategy for a generic $g_\text{eff}$ coupling to the spin. For proton and neutron couplings, $g_\text{eff}$ is rescaled by the respective nucleon contributions $\xi_P$  and $\xi_N$ to the total nuclear spin to get the final exclusions for $ g_{aPP} = g_\text{eff}/\xi_P$ and $g_{aNN} = g_\text{eff}/\xi_N $. 

The electron-coupling frequency response of the setup depends on the effect of the magnetic shield. While the effect of the shield on the ALP field itself can be neglected, the electron spins in the ferromagnetic shield ``see'' the exotic field effectively as a weak magnetic field. The shield generates an actual magnetic field in its inner volume to compensate. Therefore within the shielded volume, we have both the ALP field and the real magnetic field produced by the shield \cite{Kimball_shielding_2016}.

The frequency-dependent response of the comagnetometer to this combination of fields is discussed in Sec.\,\ref{sec:methods-electron}. Below, the Larmor precession frequency of $^3$He ($\approx 3$ Hz) the comagnetometer is, by design, insensitive to magnetic fields and its frequency response to the ALP field  \cite{padniuk_universal_2024} is the same as the nuclear frequency response but rescaled by the gyromagnetic ratio of the electron $g_{aee} = g_\text{eff}\gamma_N/\gamma_e $.

To compute upper limits on the ALP-coupling strength, the statistical properties of the technical noise distribution and the distribution of ALP DM signatures in the signal estimator $S(\omega)$ have to be considered. The ALP DM signal distribution $S_{g_\text{eff}}(\omega)$, given an effective coupling $g_\text{eff}$, is obtained by performing Monte Carlo simulations based on Eq.\,\eqref{eq:SigEstimator}. It accounts for the stochastic properties of ALP DM (Sec.\,\ref{sec:DM_model}) according to Eq.\,\eqref{eq:amplitudes}.
%takes into account the ALP DM field parameters characterized by the probability distribution from Eq.\,\eqref{eq:amplitudes},

% we take into account both the measured noise properties of $S(\omega)$ and the distribution of ALP DM signals $S_{g_\text{eff}}(\omega)$ for a given ALP-coupling strength $g_\text{eff}$ based on our assumed model of ALP DM
We use the Confidence Levels (CLs) method \cite{read_presentation_2002} to determine the limits on the coupling strength $g_\text{eff}$. 
If the measured value $S(\omega)$ is purely due to noise, it can be represented by a random number $X_n$ drawn from a non-central $\chi^2$ distribution with a scale parameter determined by the expected value based on the neighbouring points ($1/f$ fit for frequencies below 0.1\,Hz and moving average for frequencies above).
The potential contribution to $S(\omega)$ from an ALP DM signal can also be represented by a random number accounting for its stochastic nature. We compute this with 
Monte Carlo simulations and scale the ALP DM distribution with the interaction strength $g_{\textrm{eff}}$. Finally, we add the noise distribution to the signal distribution and draw from the combined distribution a random number $X_{S+n}$.

A certain value of $g_\text{eff}$ is excluded with $\text{CL}= 95\%$ at frequency $\omega$ when
\begin{equation}\label{eq:exclusioncondition}
     \frac{\mathds{P}(X_{S+n} \leq S(\omega))}{\mathds{P}(X_n \leq S(\omega))}\leq 1- \text{CL}=0.05\,,
\end{equation}
where $\mathds{P}$ indicates the probability.
The condition described by Eq.\,\eqref{eq:exclusioncondition} means that for the excluded value of $g_\text{eff}$, there is only a 5\% relative probability that an ALP DM signal contributes to the measured value of $S(\omega)$.

The sidereal frequency bin at $\omega_{\text{E}}$ is used to expand the search to frequencies below $10^{-5}$\,Hz. Due to constraints on ALP DM density  below \hbox{$m_a=10^{-22}$\,eV/$c^2$} \cite{jackson_kimball_search_2023}, 
the exclusion from the sidereal frequency bin is limited to the range from \mbox{$2.4 \times 10^{-8}$\,Hz} up to the sidereal frequency \hbox{$\omega_{\text{E}}/2\pi \approx 1.1 \times 10^{-5}$\,Hz.} In this range, the field is considered nearly constant ($\omega_a \approx 0$). The derivation of the signal model in this regime, as well as for the intermediate regime, is shown in Sec.\,\ref{sec:ultralowfrequency}.

For frequencies around $10^{-1}$\,Hz, the total measurement time is on the order of the ALP field coherence time $\tau_{\omega_a}$. The ALP oscillation is coherent for measurement times \hbox{$T \lesssim \tau_{\omega_a}/2.5$} \cite{derevianko_detecting_2018,centers_stochastic_2021}. In particular, for frequencies \hbox{$w_a/2\pi \gtrsim 5 \times 10^{-2}$\,Hz} the ALP field is not coherent over the entire measurement time ($T=92$\,days). 
Therefore, for frequencies above $5 \times 10^{-2}$\,Hz, we weaken our estimated constraints on the coupling constants by $\sqrt{T/\tau_{\omega_a}}$ to account for the incoherent averaging, as discussed in Ref.\,\cite{budker_proposal_2014}. We also consider that over several coherence times the stochastic amplitudes [$\alpha_x$, $\alpha_y$, and $\alpha_z$ in Eq.\,\eqref{eq:amplitudes}] will change. When calculating $S_{g_\text{eff}}(\omega)$ with MC simulations, we average the stochastic amplitudes over the number of coherence times during the total measurement time.

\begin{figure}[htb]    
% \label{fig:exclusion-neutron}
    \centering
    \includegraphics[width=0.49\textwidth]{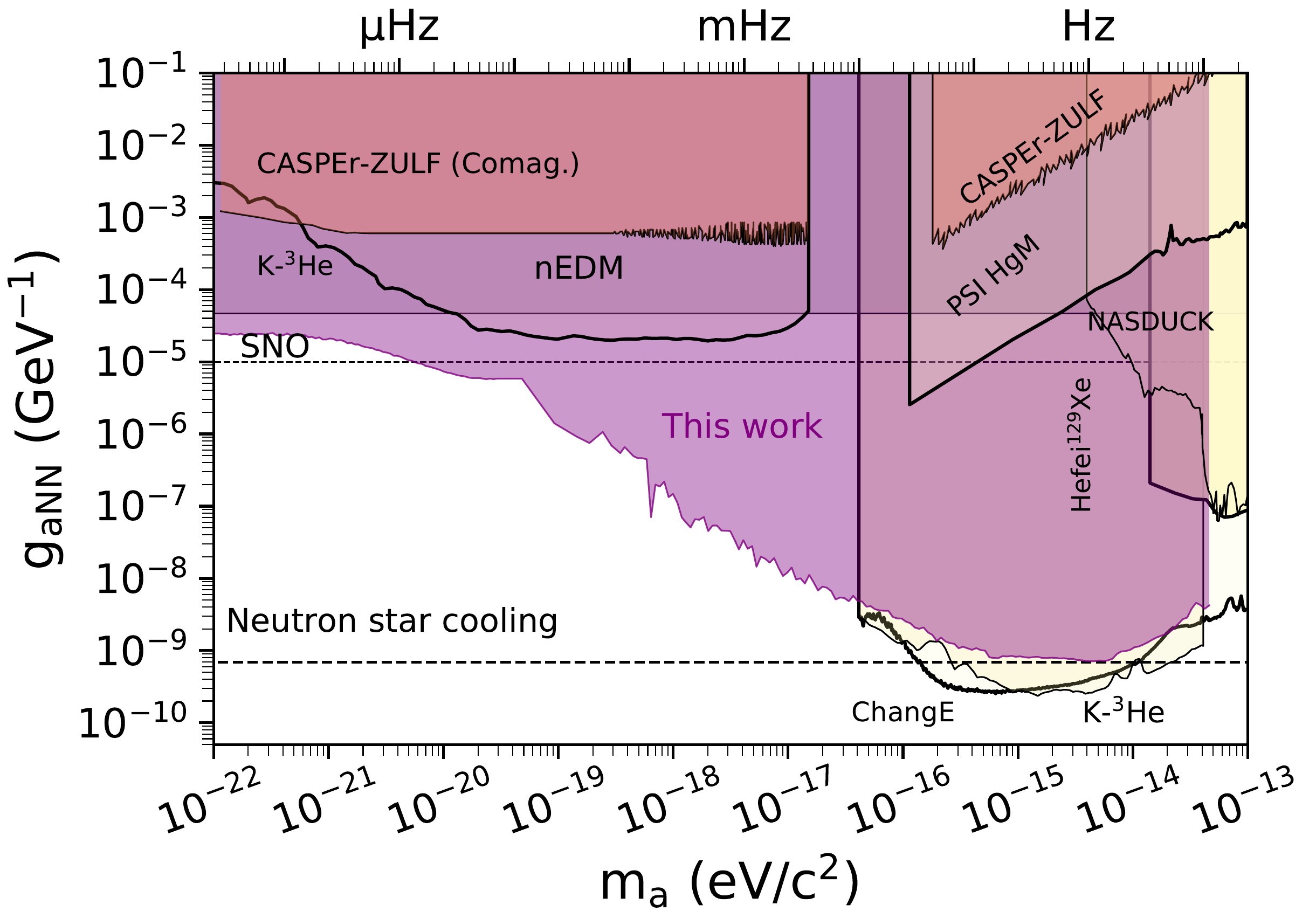}
        \caption{Exclusion plot for the neutron coupling (mean limits). Other laboratory (solid lines) and astrophysical (dashed lines) constraints are shown for reference and extracted from \cite{AxionLimits}: CASPEr-ZULF \cite{wu_search_2019,garcon_constraints_2019}, K-$^3$He \cite{lee_improved_2018,lee_laboratory_2023}, nEDM \cite{abel_search_2017}, PSI HgM \cite{abel_search_2023}, SNO \cite{bhusal_searching_2021}, NASDUCK \cite{bloch_nasduck_2022}, Hefei $^{129}$Xe \cite{jiang_search_2021}, ChangE \cite{wei_dark_2023}, and neutron-star cooling \cite{buschmann_upper_2022}.}
    \label{fig:exclusion-neutron}
\end{figure}

\begin{figure}[htb]
    \centering
    \includegraphics[width=0.49\textwidth]{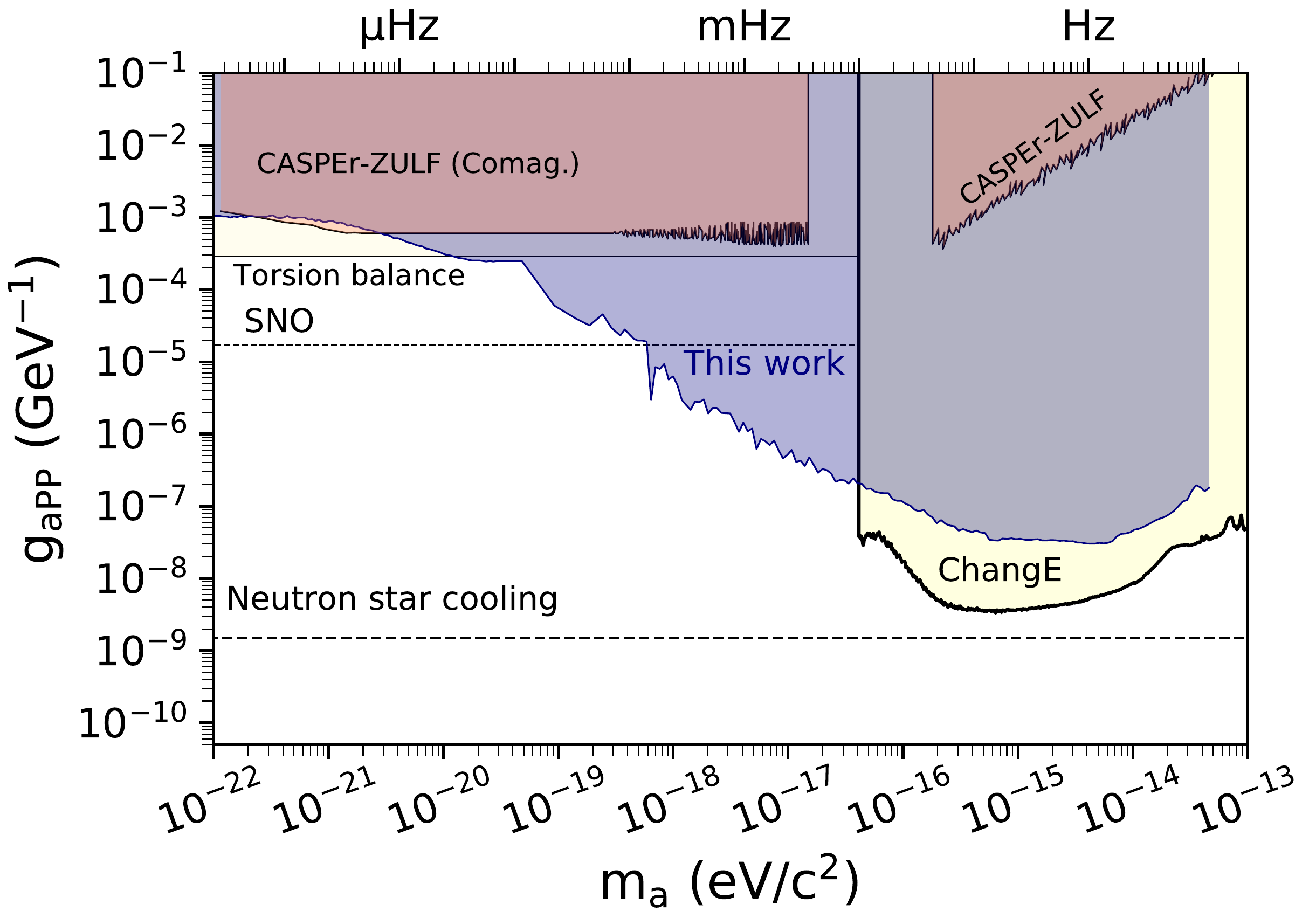}
        \caption{Exclusion plot for the proton coupling (mean limits). Other laboratory (solid lines) and astrophysical (dashed lines) constraints are shown for reference and extracted from \cite{AxionLimits}: CASPEr-ZULF \cite{wu_search_2019,garcon_constraints_2019}, ChangE \cite{wei_dark_2023}, and neutron-star cooling \cite{buschmann_upper_2022}.}
\label{fig:exclusion-proton}
\end{figure}

\begin{figure}[htb]    
% \label{fig:exclusion-neutron}
    \centering
    \includegraphics[width=0.49\textwidth]{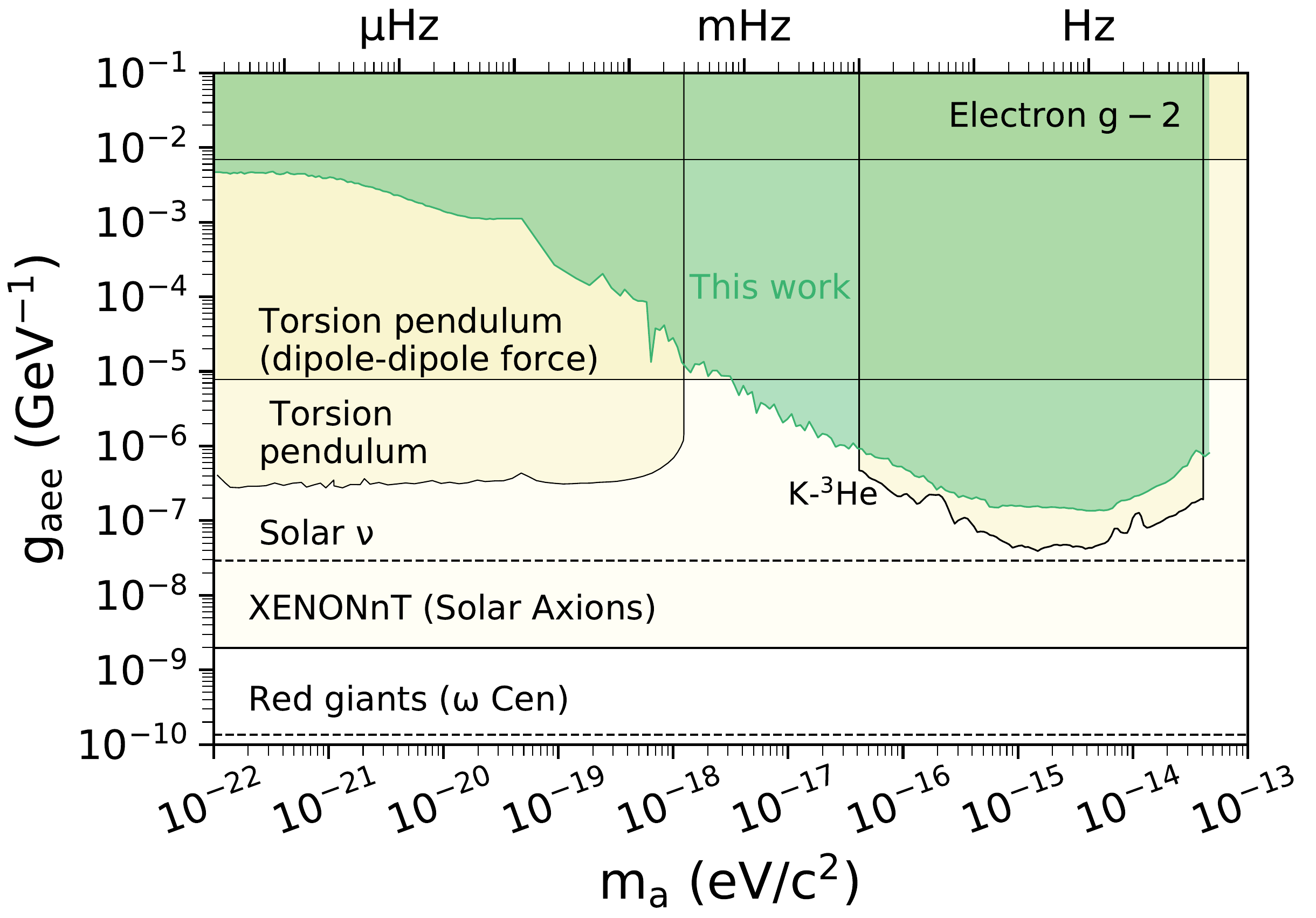}
        \caption{Exclusion plot for the electron coupling (mean limits). Other laboratory (solid lines) and astrophysical (dashed lines) constraints are shown for reference and extracted from \cite{AxionLimits}: Electron $g-2$ \cite{Yan:2019dar}, Torsion pendulum \cite{Terrano_Probe_2015, Terrano_constraints_2019}, K-$^3$He \cite{lee_laboratory_2023}, XENONnT (Solar Axions) \cite{SearchXENONnT}, Solar neutrinos \cite{Gondolo_Solar_2022}, and  red giant branch \cite{Capozzi_Axion_2020}.}
    \label{fig:exclusion-electron}
\end{figure}

The final results of this work are constraints on proton, neutron,  and electron ALP couplings $g_{aNN}$, $g_{aPP}$ and $g_{aee}$, respectively. The excluded parameter space covers a frequency range from $10^{-8}$ to $11.6$\,Hz, corresponding to masses from $10^{-22}$ to $4 \times 10^{-14}$\,eV, a total of nine orders of magnitude. Figures \ref{fig:exclusion-neutron}, \ref{fig:exclusion-proton}, and \ref{fig:exclusion-electron} show the constrained parameter space in the context of previous laboratory searches. We plot the constraints smoothed with a moving average to guide the eye (mean limits). The scatter of the constraint data is similar to that in Fig.\,\ref{fig:detectionthreshold}. In the mass range between \mbox{$1.2\times 10^{-17}$} and \mbox{$4 \times 10^{-17}$\,eV}, the exclusion improves previous constraints by 3-4 orders of magnitude in $g_{aNN}$ and $g_{aPP}$. The constraints on $g_{aee}$ improve direct DM search constraint by up to one order of magnitude and confirm the exclusions from solar axions searches and stellar physics.

Note that we do not include the results of Ref.\,\cite{bloch_axion-like_2020} in Figs.\,\ref{fig:exclusion-neutron} and \ref{fig:exclusion-electron}. Reference \cite{bloch_axion-like_2020} presented a re-analysis of the comagnetometer data from the Princeton group acquired in three different experiments \cite{vasilakis_limits_2009, brown_new_2010, lee_improved_2018}.  More recently, the Princeton group published their own re-analysis of their data \cite{lee_laboratory_2023},
%. Reference \cite{lee_laboratory_2023} 
which notes critical issues in the interpretation of their data not accounted for in Ref.\,\cite{bloch_axion-like_2020}.
We regard the Princeton analysis as the definitive interpretation of the data.

Both comagnetometers in Mainz and Krak\'ow are part of the Advanced GNOME experiment \cite{afach_what_nodate}. With additional comagnetometers currently in development, the network is set to expand, significantly enhancing its sensitivity to both ALP DM and transient events in future science runs. 

%\SP{This is not super smooth.} Other comagnetometers are in development and will join future science runs, allowing for an increased sensitivity to ALP DM as well as to transient events. Transient event searches greatly benefit from a network of more than two sensors.

At the same time, this experiment is part of the CASPEr family of experiments \cite{garcon_constraints_2019,wu_search_2019}, significantly improving previous CASPEr results (up to seven orders of magnitude) while also extending the covered mass range, see Figs.\,\ref{fig:exclusion-neutron} and \ref{fig:exclusion-proton}.


% 575 words
\section{Methods}\label{sec:methods}

%\begin{itemize}
    %\item Short and precise, no plots just references \\\EK{Manu: Hold my beer}
%\end{itemize}
\subsection{Experimental setup}\label{sec:methods-setup}
\begin{figure}[htb]
    \centering
    \includegraphics[width=0.5\textwidth]{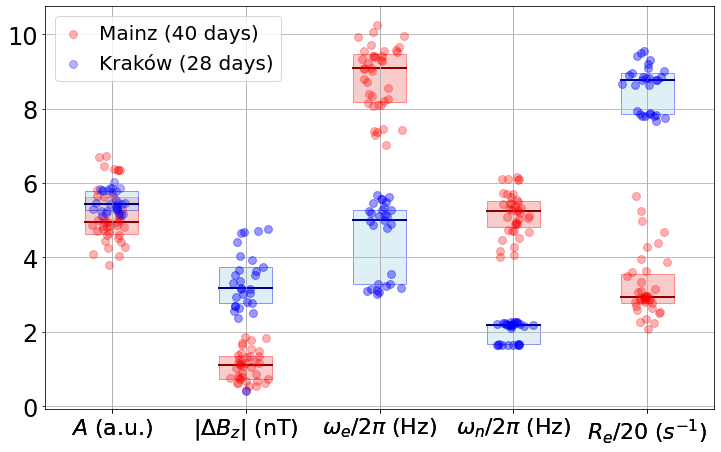}
        \caption{Summary of the frequency-response-fit results used for sensor calibration in Mainz and Krak\'ow: response function amplitude $A$, detuning from the compensation point $\Delta B_z$, alkali-metal and noble gas Larmor frequencies $\omega_e$ and $\omega_n$, alkali-metal polarization relaxation rate $R_e$ (divided by 20 only for the better visualization with other parameters). The fit parameters are described in detail in Ref. \cite{padniuk_universal_2024}. Horizontal bars designate the median values of parameters and the shaded regions extend to the first quartiles.}
    \label{fig:freqresponse}
\end{figure}

The interferometer is composed of two self-compensating comagnetometers located about 1000\,km apart: one in Mainz, Germany and the other in Krak\'ow, Poland.
The two self-compensating comagnetometers are similar to that reported in Refs.\,\cite{klingerPRA2023,padniuk_universal_2024}. At the core of the Mainz (Krak\'ow) comagnetometer system is a spherical cell heated to about $180^\circ$C and mounted inside a four-layer magnetic shield. The cell is filled with 3\,amg of $^3$He and 50\,Torr of N$_2$ and loaded with a drop of an alkali-metal mixture with 1\%\,$^{87}$Rb and 99\% natural-abundance K (molar fractions). Spins are optically pumped with a $30$\,mW/cm$^2$ ($50$\,mW/cm$^2$) circularly-polarized light tuned to the center of the Rb D$_1$ (D$_2$) line. The readout is realized by monitoring the polarization rotation of a $\sim 15$\,mW/cm$^2$ ($1$\,mW/cm$^2$) linearly-polarized light detuned about 0.5\,nm from the K D$_1$ line. 
To reduce the influence of the magnetic-field noise at low frequencies, the comagnetometers are operated in the self-compensating regime \cite{kornack_dynamics_2002}. To operate in this regime, a $B_z$ (compensation) field of about 100\,nT (50\,nT) is applied in Mainz (Krak\'ow). In the Mainz comagnetometer, we modulate the $B_x$ field with a 80\,Hz sine wave. The signal demodulated at that frequency exhibits a resonance that is used to lock the system to the compensation point and therefore follows slow drifts of the equilibrium compensation field \cite{klingerPRA2023}. 
 The sensitivity of both comagnetometers to exotic nucleon couplings is estimated with a daily (every 25\,h) calibration pulse according to the calibration procedure described in Ref.\,\cite{padniuk_universal_2024}. In the Krak\'ow comagnetometer, the drifts from the compensation point are corrected depending on demand after applying the calibration pulse (see, for example, two groups of fitted values of $\omega_e$ and $\omega_n$ in Fig.\, \ref{fig:freqresponse}, corresponding to two time-separated datasets), while there is no such a correction procedure in Mainz. The response to the calibration pulse is fitted with such parameters as the amplitude of the response that relates the voltage output to pseudo-magnetic fields, the detuning from the compensation point $\Delta B_z$, the Larmor frequency of both electron spin $\omega_{e}$, and $^3$He nuclear spin $\omega_n$ and the relaxation rate of electrons $R_e$ (at the current stage of the work, nuclear relaxation $R_n$ is ignored). Figure\,\ref{fig:freqresponse} summarizes the fit parameters for all 25 h datasets. As shown, there is little variance in the parameters during the whole run. The detuning from the compensation point $\Delta B_z$ remains below 5\,nT, which is within 5\,\% (10\,\%)  of to the leading field $\sim100$\,nT ($\sim 50$\,nT) in Mainz (Krak\'ow).
% units for amplitude and R_e, make points larger

\subsection{Weights of the ALP signal estimator $S(\omega)$}
\label{sec:methods-weights}
 The weights used in the signal estimator [see Eqs.\,\eqref{eq:sideband_interference} and\,\eqref{eq:SigEstimator}] are defined in the following way
\begin{equation}
\begin{split}
    a^M_{\pm}&= \frac{\sin\theta_M}{(\sigma^M_{A_\pm })^2}\, ,\\
        a^K_{\pm}&= \frac{\sin\theta_K}{(\sigma^K_{A_\pm })^2}\, , \\
        b^M&=\frac{2\cos\theta_M}{(\sigma^M_{|A|^2})^2} \approx 0\, ,\\
        b^K&=\frac{2\cos\theta_K}{(\sigma^K_{|A|^2})^2}\, ,\\
        b_\pm^{K+M}&=\frac{2}{(\sigma^{K+M}_{|A_\pm|^2})^2}\, ,\\
    % (a_{\text{sideband},k})= \frac{P_{\omega_{rot} \perp k}P_{\omega_{rot} \perp v_e}}{\sigma^2_{k,i}} e^{i\phi} \, ,
\end{split}
\end{equation}
where $a^i_\pm$ and $b^i_\pm$ are the weights with $\pm$ designating the higher ($+$) and lower ($-$) frequency sideband and index \hbox{$i=M,K,K+M$} indicates the Mainz, Krak\'ow, and combined signal, respectively.
The $\sigma_A$ ($\sigma_{|A|^2}$) represents the standard deviation of the Mainz and Krak\'ow amplitudes (power) in the frequency bin of interest. For the interfered sidebands, the weights are resulting from error propagation of $b^{K+M}_{\pm}$ according to Eq.\,\eqref{eq:sideband_interference}. The factor 2 in $b^M$, $b^K$ and $b^{K+M}_{\pm}$ is given by the expected ALP DM signal in the carrier being twice the sidebands [Eq.\,\eqref{eq:amplitudes}]. 

\subsection{Noise distribution and signal insertion demonstration}\label{sec:methods:noise}

\begin{figure}[htb]
    \centering
    \includegraphics[width=0.5\textwidth]{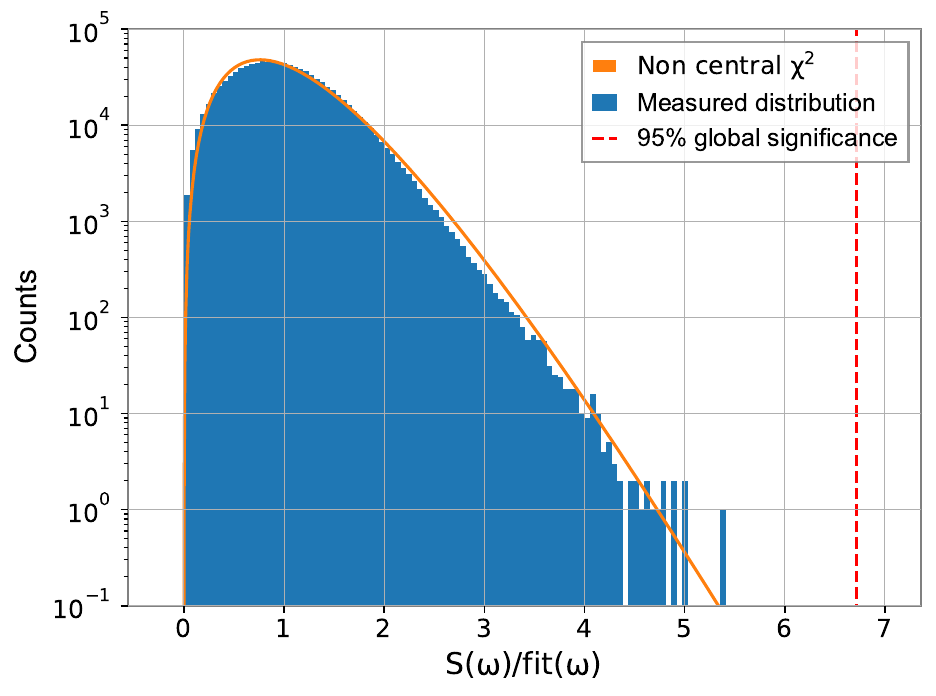}
        \caption{Histogram of the normalized signal estimator $S(\omega)/\text{fit}(\omega)$, obtained by dividing the signal estimator $S(\omega)$ to the expected noise, fit$(\omega)$. The global threshold at each frequency is determined by fitting the normalized signal estimator with a non-central $\chi^2$ distribution with six degrees of freedom. This enables the calculation of the limits according to Eq.\,\eqref{eq:exclusioncondition}. Note that the $y$-axis is in logarithmic scale to show the tail of the distribution in more details.}
    \label{fig:noncentraldistro}
\end{figure}

In order to derive the global threshold and the limits, it is important to understand the expected distribution of the signal estimator $S(\omega)$. The signal power estimator $S^2(\omega)$ is a sum of six squares of normally-distributed numbers: the real and imaginary Fourier coefficients from three frequency bins. However, to compare noise at different frequencies, we normalize (``whiten'') each bin by the estimated mean noise, $\text{fit}(\omega)$. Note that the expected value of the noise is proportional to the standard deviation of $S(\omega)$, namely $\Delta S(\omega)$. 
%The distribution of interest is given by the signal excess $S(\omega)/\Delta S(\omega)$ and follows a non-central $\chi^2$-distribution.
The signal estimator excess square $S^2(\omega)/\Delta S^2(\omega)$ follows a non-central $\chi^2$-distribution, defined as $Z=\sum_{n=1}^N X_n^2$, where  $\{X_n\}$ is a set of normally distributed random numbers with different means and same variance. From propagation of errors, we have that $\Delta S^2(\omega) = 2S(\omega)\Delta S(\omega)$. As $S(\omega)/\Delta S(\omega)= 2S^2(\omega)/\Delta S^2(\omega)$, they both follow the same distribution.
This is also the case for $S(\omega)/\text{fit}(\omega) \sim S(\omega)/\Delta S(\omega)$, which is the estimator used to calculate the global threshold and the limits. The fitted non-central $\chi^2$ distribution reproduces the histogram of $S(\omega)/\text{fit}(\omega)$, especially its tail, which is critical for
claiming detection, see Fig.\,\ref{fig:noncentraldistro}.

Let us now consider the modification of the distribution induced by an ALP DM signal. Both noise and ALP DM signal are expected to have the same distribution: Gaussian variables in both Fourier quadratures of the raw Fourier spectrum (before combining the Mainz and Krak\'ow carrier and sidebands). We inject the ALP DM signal in the raw Fourier spectrum as a normally-distributed random variable and then the data are combined to obtain the signal estimator $S(\omega)$ [see Eqs.\,\eqref{eq:sideband_interference} and\,\eqref{eq:SigEstimator}].

\begin{figure}[htb]
    \centering
    \includegraphics[width=0.5\textwidth]{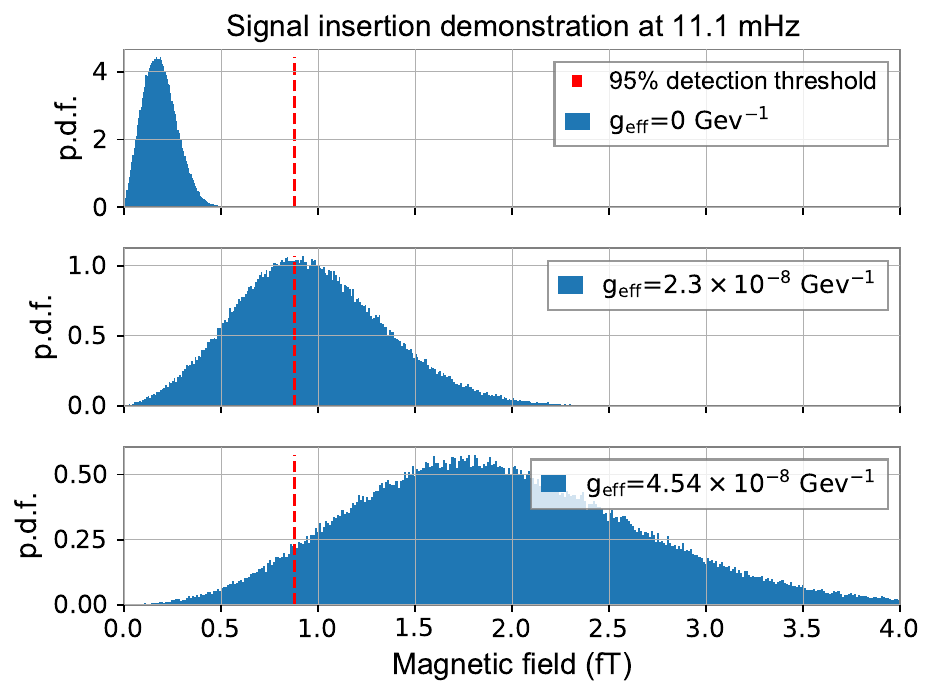}
        \caption{Insertion of a signal at a frequency bin situated at around 11.1\,mHz. The shape of both noise and signal are the same, since they come from the same distribution. By changing the inserted effective coupling $g_\text{eff}$, $S(\omega)$ reaches above the detection threshold at this frequency. Due to the measurement time being shorter than the coherence time, this distribution is only sampled once, so it is possible that even for a large coupling strength the amplitude might still be below the threshold.}
    \label{fig:insertion}
\end{figure}

Figure\,\ref{fig:insertion} shows the distribution of $S(\omega/2\pi= 11.1$\,mHz) with an injected ALP DM signal for different coupling strengths $g_\text{eff}$ computed according to Eq.\,\eqref{eq:amplitudes}. To sample the stochastic ALP parameter space, we injected $10^6$ different sets of random amplitudes $\alpha_i$ and phases $\phi_i$. The resulting distributions are the sum of the noise and ALP signal distributions. As shown, increasing $g_\text{eff}$ increases the mean and the variance of the distribution, which has to be taken into account when setting limits.

\subsection{Derivation of the ALP signal in the ultra-low oscillating regime $\omega_a \ll \omega_{\text{E}}$} \label{sec:ultralowfrequency}

%Here, we present a derivation of the expected signal model for $\omega_a \ll \omega_{\text{E}}$, as well as for the intermediate regime.
Consider a single station and a quasi-constant oscillating ALP field at $\omega_a \ll \omega_{\text{E}}$. The expected signal, defined in Eq.\,\eqref{eq:modulated_signal}, becomes (ignoring the constant terms) 
\begin{equation}
\label{eq:low_frequency_signal}
\begin{split}
         \lim_{\omega_a \rightarrow 0}\boldsymbol{\nabla} a(t) \cdot \boldsymbol{\hat{m}}(t)   =  \sin\theta \Big\lbrace &\alpha_x \cos\phi_x \sin(\omega_{\text{E}} t + \phi_{\text{E}}) \\ 
         +&    \alpha_y \cos\phi_y \cos(\omega_{\text{E}} t + \phi_{\text{E}})\Big\rbrace \,. 
\end{split}
\end{equation} 
Therefore, the oscillating signal expected in the sensor is driven only by the rotation of Earth at $\omega_{\text{E}}$ with an amplitude 
\begin{equation}
\label{eq:amplitude_sidereal}
    A_{\omega_{\text{E}}} = g_\text{eff}\sin\theta ~\sqrt{(\alpha_x \cos\phi_x)^2 +(\alpha_y \cos\phi_y)^2}\,,
\end{equation}

In contrast to the case of $\omega_a > \omega_{\text{E}}$ [Eq.~\eqref{eq:ALP-power}], the ALP DM signal amplitude explicitly depends on the random phases $\phi_x$ and $\phi_y$. In this regime, where $\omega_a \approx 0$, the sensors sample less than a period of the ALP DM field; and there is a probability of measuring near a zero crossing of the oscillation (when $A_{\omega_{\text{E}}} \approx 0$). To account for this and the intermediate regime, where some of the ALP DM oscillation is still measured, we perform
Monte Carlo simulations of the integrated power of the respective oscillation fractions. We define this factor as
\begin{equation}
    \kappa(\omega_a)=\int_0^T dt\frac{ \alpha^2_x\cos^2(\omega_a t+\phi_x) + \alpha^2_y\cos^2(\omega_a t+\phi_y)}{T(\alpha_x^2+\alpha_y^2)/2}\,
\end{equation} for uniformly distributed phases $\phi_x$ and $\phi_y$ and the respective $\omega_a$. $T=92$\,days is the total measurement time. The constraints at a frequency $\omega_a<\omega_\text{E}$ are given by the product of $\kappa(\omega_a)$ and the constraint at $\omega_{\text{E}}$.

\subsection{Comagnetometer response to exotic electron interaction}\label{sec:methods-electron}

%In this subsection, we discuss the response of the comagnetometer to exotic electron interactions such as ALP-electron interactions. 

As explicitly shown in Ref.\,\cite{padniuk_universal_2024}, the direct coupling of the ALP DM with electrons results in a frequency response that is significantly different from those of neutrons and protons. To accurately estimate the response of the system, one has to consider all possible manifestations of the ALP-electron interaction. The comagnetometer vapor cell is surrounded by a mu-metal shield that cancels the external magnetic fields. However, due to the electron-based mechanism of magnetic shielding, the mu-metal also responds to an exotic electron interaction and induces a ``compensating'' magnetic field inside the shield (the response of the shield in the case of an exotic electron perturbation has been discussed in Ref.\,\cite{Kimball_shielding_2016}). For the purpose of searching for ALP-electron coupling in the analyzed frequency range, we assume that the ALP-electron shielding factor is of the order of the magnetic shielding factor.
%However, in addition to the spin contribution in the shield, there is also a (relatively small) orbital contribution that would not couple to an ALP field \cite{Khamis:2024oqa}. This results in a (relatively small) uncertainty in the compensating field generated by the shield. 

 The ALP-electron interaction for a comagnetometer and a mu-metal shield manifests in the following way. The shield generates an opposing real magnetic field as a response to the applied exotic effective field at the sensor position. For a comagnetometer inside the shield, the two fields are superimposed. The electron is influenced by both opposing (magnetic and exotic) fields such that the perturbation of the electron, and consequently its response, is canceled. However, the magnetic field generated by the shield is picked up by the $^3$He nuclear spins, which do not directly respond to the ALP electron interaction. Hence, for 100\% shielding (i.e., when the induced magnetic field is equal and opposite to the effective exotic field), the frequency response has the same shape as the comagnetometer response to exotic nuclear spin interactions. It is rescaled by the response of the electrons in the shield to the ALP interaction, which is assumed to be flat in the frequency range of interest, effectively attenuating the response by factor $\gamma_N/\gamma_e$. This is the scenario we assume when setting the limits in Fig.\,\ref{fig:exclusion-electron}.

\begin{figure}[htb]    
    \centering
    \includegraphics[width=0.48\textwidth]{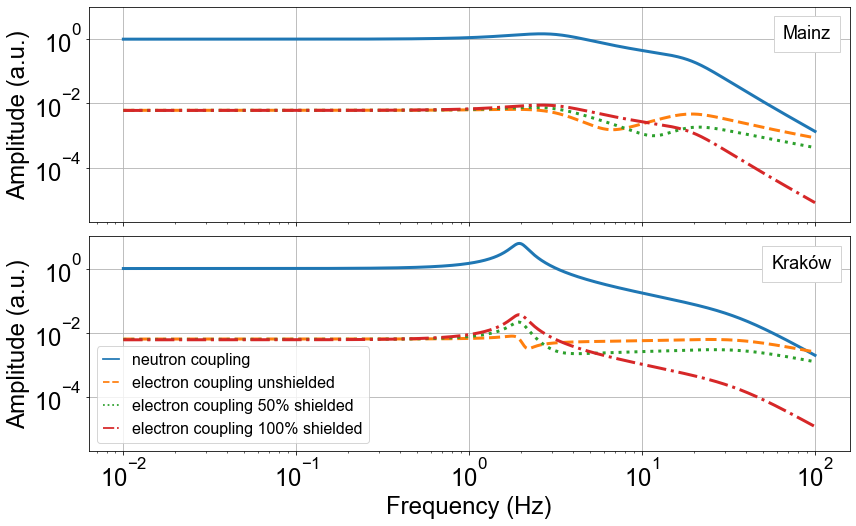}
        \caption{Comagnetometer frequency response to exotic electronic perturbations in different shielding scenarios in Mainz (top) and Krak\'ow (bottom): unshielded (dashed), 50\%  shielded (dotted), fully shielded (dash-dotted). The neutron coupling frequency response (full line) is shown for reference.} 
        \label{fig:electron-response}
\end{figure}

In general, the nuclear and electron spin perturbation in the comagnetometer is different for arbitrary shielding factors and then the self-compensating mechanism would not apply. However, an interesting case is when the shield responds only at a 50\% level to the exotic electron coupling. In the framework of Ref.\,\cite{padniuk_response_2022}, this corresponds to $\alpha_e=0.5$ and $\alpha_n=-0.5$. In this case, although the magnetic and exotic interactions are equal in strength, they have opposing directions and hence the comagnetometer remains sensitive to the electron exotic interaction. The shape of the frequency response is a combination of the electron and nuclear spin responses. 

Figure\,\ref{fig:electron-response} shows the comagnetometer response for three different shielding scenarios. They converge at frequencies below the nuclear Larmor frequency ($\sim3$\,Hz), since the nuclear spin dominates the dynamics in this regime. Thus, the limits on ALP-electron coupling of this work in this frequency range are almost invariant for any shielding. However, for searches beyond 11.6\,Hz, the response changes significantly. For high frequencies ($\sim100$\,Hz), the unshielded electron response is greater compared to the fully shielded response and a detailed shielding model should be considered while setting limits. For reference, the nuclear spin response to exotic neutron interactions is also displayed in Fig.\,\ref{fig:electron-response}.

\section{Acknowledgments}
We would like to acknowledge  Gary Centers, Wolfgang Gradl, Julian Walter, and Yuzhe Zang for helpful discussion. We acknowledge support by the German Research Foundation (DFG) within the German Excellence Strategy (Project ID 39083149); by work from COST Action COSMIC WISPers CA21106, supported by COST (European Cooperation in Science and Technology). SP, MP, and MS acknowledge the support from the National Science Center, Poland within the OPUS program (2020/39/B/ST2/01524) and GL acknowledges support from the Excellence
Initiative – Research University of the Jagiellonian University in Krak\'ow.
The work of DFJK was supported by the U.S. National Science Foundation under grant PHYS-2110388. The work of AOS was supported by the U.S. National Science Foundation CAREER grant PHY-2145162, and by the Gordon and Betty Moore Foundation, grant DOI 10.37807/gbmf12248.

\newpage

%\bibliography{Zotero_ref, BIB, Fluffballs_Zotero,}
\bibliography{Bibliography.bib}

\begin{thebibliography}{62}%
\makeatletter
\providecommand \@ifxundefined [1]{%
 \@ifx{#1\undefined}
}%
\providecommand \@ifnum [1]{%
 \ifnum #1\expandafter \@firstoftwo
 \else \expandafter \@secondoftwo
 \fi
}%
\providecommand \@ifx [1]{%
 \ifx #1\expandafter \@firstoftwo
 \else \expandafter \@secondoftwo
 \fi
}%
\providecommand \natexlab [1]{#1}%
\providecommand \enquote  [1]{``#1''}%
\providecommand \bibnamefont  [1]{#1}%
\providecommand \bibfnamefont [1]{#1}%
\providecommand \citenamefont [1]{#1}%
\providecommand \href@noop [0]{\@secondoftwo}%
\providecommand \href [0]{\begingroup \@sanitize@url \@href}%
\providecommand \@href[1]{\@@startlink{#1}\@@href}%
\providecommand \@@href[1]{\endgroup#1\@@endlink}%
\providecommand \@sanitize@url [0]{\catcode `\\12\catcode `\$12\catcode
  `\&12\catcode `\#12\catcode `\^12\catcode `\_12\catcode `\%12\relax}%
\providecommand \@@startlink[1]{}%
\providecommand \@@endlink[0]{}%
\providecommand \url  [0]{\begingroup\@sanitize@url \@url }%
\providecommand \@url [1]{\endgroup\@href {#1}{\urlprefix }}%
\providecommand \urlprefix  [0]{URL }%
\providecommand \Eprint [0]{\href }%
\providecommand \doibase [0]{https://doi.org/}%
\providecommand \selectlanguage [0]{\@gobble}%
\providecommand \bibinfo  [0]{\@secondoftwo}%
\providecommand \bibfield  [0]{\@secondoftwo}%
\providecommand \translation [1]{[#1]}%
\providecommand \BibitemOpen [0]{}%
\providecommand \bibitemStop [0]{}%
\providecommand \bibitemNoStop [0]{.\EOS\space}%
\providecommand \EOS [0]{\spacefactor3000\relax}%
\providecommand \BibitemShut  [1]{\csname bibitem#1\endcsname}%
\let\auto@bib@innerbib\@empty
%</preamble>
\bibitem [{\citenamefont {Bertone}\ \emph {et~al.}(2005)\citenamefont
  {Bertone}, \citenamefont {Hooper},\ and\ \citenamefont
  {Silk}}]{Bertone2005Jan}%
  \BibitemOpen
  \bibfield  {author} {\bibinfo {author} {\bibfnamefont {G.}~\bibnamefont
  {Bertone}}, \bibinfo {author} {\bibfnamefont {D.}~\bibnamefont {Hooper}},\
  and\ \bibinfo {author} {\bibfnamefont {J.}~\bibnamefont {Silk}},\ }\bibfield
  {title} {\bibinfo {title} {{Particle dark matter: evidence, candidates and
  constraints}},\ }\href {https://doi.org/10.1016/j.physrep.2004.08.031}
  {\bibfield  {journal} {\bibinfo  {journal} {Phys. Rep.}\ }\textbf {\bibinfo
  {volume} {405}},\ \bibinfo {pages} {279} (\bibinfo {year}
  {2005})}\BibitemShut {NoStop}%
\bibitem [{\citenamefont {Feng}(2010)}]{Feng2010Sep}%
  \BibitemOpen
  \bibfield  {author} {\bibinfo {author} {\bibfnamefont {J.~L.}\ \bibnamefont
  {Feng}},\ }\bibfield  {title} {\bibinfo {title} {{Dark Matter Candidates from
  Particle Physics and Methods of Detection}},\ }\href
  {https://doi.org/10.1146/annurev-astro-082708-101659} {\bibfield  {journal}
  {\bibinfo  {journal} {Annu. Rev. Astron. Astrophys.}\ }\textbf {\bibinfo
  {volume} {48}},\ \bibinfo {pages} {495} (\bibinfo {year} {2010})}\BibitemShut
  {NoStop}%
\bibitem [{\citenamefont {Gorenstein}\ and\ \citenamefont
  {Tucker}(2014)}]{Gorenstein2014Jun}%
  \BibitemOpen
  \bibfield  {author} {\bibinfo {author} {\bibfnamefont {P.}~\bibnamefont
  {Gorenstein}}\ and\ \bibinfo {author} {\bibfnamefont {W.}~\bibnamefont
  {Tucker}},\ }\bibfield  {title} {\bibinfo {title} {{Astronomical Signatures
  of Dark Matter}},\ }\href {https://doi.org/10.1155/2014/878203} {\bibfield
  {journal} {\bibinfo  {journal} {Adv. High Energy Phys.}\ }\textbf {\bibinfo
  {volume} {2014}},\ \bibinfo {pages} {878203} (\bibinfo {year}
  {2014})}\BibitemShut {NoStop}%
\bibitem [{\citenamefont {Antypas}\ \emph {et~al.}(2022)\citenamefont
  {Antypas}, \citenamefont {Banerjee}, \citenamefont {Bartram}, \citenamefont
  {Baryakhtar}, \citenamefont {Betz} \emph {et~al.}}]{Antypas2022Mar}%
  \BibitemOpen
  \bibfield  {author} {\bibinfo {author} {\bibfnamefont {D.}~\bibnamefont
  {Antypas}}, \bibinfo {author} {\bibfnamefont {A.}~\bibnamefont {Banerjee}},
  \bibinfo {author} {\bibfnamefont {C.}~\bibnamefont {Bartram}}, \bibinfo
  {author} {\bibfnamefont {M.}~\bibnamefont {Baryakhtar}}, \bibinfo {author}
  {\bibfnamefont {J.}~\bibnamefont {Betz}}, \emph {et~al.},\ }\bibfield
  {title} {\bibinfo {title} {{New Horizons: Scalar and Vector Ultralight Dark
  Matter}},\ }\bibfield  {journal} {\bibinfo  {journal} {arXiv}\ }\href
  {https://doi.org/10.48550/arXiv.2203.14915} {10.48550/arXiv.2203.14915}
  (\bibinfo {year} {2022}),\ \Eprint {https://arxiv.org/abs/2203.14915}
  {2203.14915} \BibitemShut {NoStop}%
\bibitem [{\citenamefont {Abbott}\ and\ \citenamefont
  {Sikivie}(1983)}]{abbott_cosmological_1983}%
  \BibitemOpen
  \bibfield  {author} {\bibinfo {author} {\bibfnamefont {L.}~\bibnamefont
  {Abbott}}\ and\ \bibinfo {author} {\bibfnamefont {P.}~\bibnamefont
  {Sikivie}},\ }\bibfield  {title} {\bibinfo {title} {A cosmological bound on
  the invisible axion},\ }\href
  {https://doi.org/https://doi.org/10.1016/0370-2693(83)90638-X} {\bibfield
  {journal} {\bibinfo  {journal} {Physics Letters B}\ }\textbf {\bibinfo
  {volume} {120}},\ \bibinfo {pages} {133} (\bibinfo {year}
  {1983})}\BibitemShut {NoStop}%
\bibitem [{\citenamefont {Preskill}\ \emph {et~al.}(1983)\citenamefont
  {Preskill}, \citenamefont {Wise},\ and\ \citenamefont
  {Wilczek}}]{preskill_cosmology_1983}%
  \BibitemOpen
  \bibfield  {author} {\bibinfo {author} {\bibfnamefont {J.}~\bibnamefont
  {Preskill}}, \bibinfo {author} {\bibfnamefont {M.~B.}\ \bibnamefont {Wise}},\
  and\ \bibinfo {author} {\bibfnamefont {F.}~\bibnamefont {Wilczek}},\
  }\bibfield  {title} {\bibinfo {title} {Cosmology of the invisible axion},\
  }\href {https://doi.org/https://doi.org/10.1016/0370-2693(83)90637-8}
  {\bibfield  {journal} {\bibinfo  {journal} {Phys. Lett. B}\ }\textbf
  {\bibinfo {volume} {120}},\ \bibinfo {pages} {127} (\bibinfo {year}
  {1983})}\BibitemShut {NoStop}%
\bibitem [{\citenamefont {Dine}\ and\ \citenamefont
  {Fischler}(1983)}]{dine_not-so-harmless_1983}%
  \BibitemOpen
  \bibfield  {author} {\bibinfo {author} {\bibfnamefont {M.}~\bibnamefont
  {Dine}}\ and\ \bibinfo {author} {\bibfnamefont {W.}~\bibnamefont
  {Fischler}},\ }\bibfield  {title} {\bibinfo {title} {The not-so-harmless
  axion},\ }\href
  {https://doi.org/https://doi.org/10.1016/0370-2693(83)90639-1} {\bibfield
  {journal} {\bibinfo  {journal} {Phys. Lett. B}\ }\textbf {\bibinfo {volume}
  {120}},\ \bibinfo {pages} {137} (\bibinfo {year} {1983})}\BibitemShut
  {NoStop}%
\bibitem [{\citenamefont {Co}\ \emph {et~al.}(2020)\citenamefont {Co},
  \citenamefont {Hall},\ and\ \citenamefont {Harigaya}}]{co_axion_2020}%
  \BibitemOpen
  \bibfield  {author} {\bibinfo {author} {\bibfnamefont {R.~T.}\ \bibnamefont
  {Co}}, \bibinfo {author} {\bibfnamefont {L.~J.}\ \bibnamefont {Hall}},\ and\
  \bibinfo {author} {\bibfnamefont {K.}~\bibnamefont {Harigaya}},\ }\bibfield
  {title} {\bibinfo {title} {Axion {Kinetic} {Misalignment} {Mechanism}},\
  }\href {https://doi.org/10.1103/PhysRevLett.124.251802} {\bibfield  {journal}
  {\bibinfo  {journal} {Phys. Rev. Lett.}\ }\textbf {\bibinfo {volume} {124}},\
  \bibinfo {pages} {251802} (\bibinfo {year} {2020})}\BibitemShut {NoStop}%
\bibitem [{\citenamefont {Marsh}(2016)}]{Marsh2016Jul}%
  \BibitemOpen
  \bibfield  {author} {\bibinfo {author} {\bibfnamefont {D.~J.~E.}\
  \bibnamefont {Marsh}},\ }\bibfield  {title} {\bibinfo {title} {{Axion
  cosmology}},\ }\href {https://doi.org/10.1016/j.physrep.2016.06.005}
  {\bibfield  {journal} {\bibinfo  {journal} {Phys. Rep.}\ }\textbf {\bibinfo
  {volume} {643}},\ \bibinfo {pages} {1} (\bibinfo {year} {2016})}\BibitemShut
  {NoStop}%
\bibitem [{\citenamefont {Graham}\ and\ \citenamefont
  {Rajendran}(2013)}]{Graham2013Aug}%
  \BibitemOpen
  \bibfield  {author} {\bibinfo {author} {\bibfnamefont {P.~W.}\ \bibnamefont
  {Graham}}\ and\ \bibinfo {author} {\bibfnamefont {S.}~\bibnamefont
  {Rajendran}},\ }\bibfield  {title} {\bibinfo {title} {{New observables for
  direct detection of axion dark matter}},\ }\href
  {https://doi.org/10.1103/PhysRevD.88.035023} {\bibfield  {journal} {\bibinfo
  {journal} {Phys. Rev. D}\ }\textbf {\bibinfo {volume} {88}},\ \bibinfo
  {pages} {035023} (\bibinfo {year} {2013})}\BibitemShut {NoStop}%
\bibitem [{\citenamefont {Budker}\ \emph {et~al.}(2014)\citenamefont {Budker},
  \citenamefont {Graham}, \citenamefont {Ledbetter}, \citenamefont
  {Rajendran},\ and\ \citenamefont {Sushkov}}]{budker_proposal_2014}%
  \BibitemOpen
  \bibfield  {author} {\bibinfo {author} {\bibfnamefont {D.}~\bibnamefont
  {Budker}}, \bibinfo {author} {\bibfnamefont {P.~W.}\ \bibnamefont {Graham}},
  \bibinfo {author} {\bibfnamefont {M.}~\bibnamefont {Ledbetter}}, \bibinfo
  {author} {\bibfnamefont {S.}~\bibnamefont {Rajendran}},\ and\ \bibinfo
  {author} {\bibfnamefont {A.~O.}\ \bibnamefont {Sushkov}},\ }\bibfield
  {title} {\bibinfo {title} {Proposal for a {Cosmic} {Axion} {Spin}
  {Precession} {Experiment} ({CASPEr})},\ }\href
  {https://doi.org/10.1103/PhysRevX.4.021030} {\bibfield  {journal} {\bibinfo
  {journal} {Phys. Rev. X}\ }\textbf {\bibinfo {volume} {4}},\ \bibinfo {pages}
  {021030} (\bibinfo {year} {2014})}\BibitemShut {NoStop}%
\bibitem [{\citenamefont {Jackson~Kimball}\ and\ \citenamefont
  {Van~Bibber}(2023)}]{jackson_kimball_search_2023}%
  \BibitemOpen
  \bibinfo {editor} {\bibfnamefont {D.~F.}\ \bibnamefont {Jackson~Kimball}}\
  and\ \bibinfo {editor} {\bibfnamefont {K.}~\bibnamefont {Van~Bibber}},\
  eds.,\ \href {https://doi.org/10.1007/978-3-030-95852-7} {\emph {\bibinfo
  {title} {The {Search} for {Ultralight} {Bosonic} {Dark} {Matter}}}}\
  (\bibinfo  {publisher} {Springer International Publishing},\ \bibinfo
  {address} {Cham},\ \bibinfo {year} {2023})\BibitemShut {NoStop}%
\bibitem [{\citenamefont {Graham}\ \emph {et~al.}(2015)\citenamefont {Graham},
  \citenamefont {Irastorza}, \citenamefont {Lamoreaux}, \citenamefont
  {Lindner},\ and\ \citenamefont {Bibber}}]{graham_experimental_2015}%
  \BibitemOpen
  \bibfield  {author} {\bibinfo {author} {\bibfnamefont {P.~W.}\ \bibnamefont
  {Graham}}, \bibinfo {author} {\bibfnamefont {I.~G.}\ \bibnamefont
  {Irastorza}}, \bibinfo {author} {\bibfnamefont {S.~K.}\ \bibnamefont
  {Lamoreaux}}, \bibinfo {author} {\bibfnamefont {A.}~\bibnamefont {Lindner}},\
  and\ \bibinfo {author} {\bibfnamefont {K.~A.~v.}\ \bibnamefont {Bibber}},\
  }\bibfield  {title} {\bibinfo {title} {Experimental {Searches} for the
  {Axion} and {Axion}-{Like} {Particles}},\ }\href
  {https://doi.org/10.1146/annurev-nucl-102014-022120} {\bibfield  {journal}
  {\bibinfo  {journal} {Annu. Rev. Nucl. Part. Sci.}\ }\textbf {\bibinfo
  {volume} {65}},\ \bibinfo {pages} {485} (\bibinfo {year} {2015})},\ \bibinfo
  {note} {publisher: Annual Reviews}\BibitemShut {NoStop}%
\bibitem [{\citenamefont {Foster}\ \emph {et~al.}(2021)\citenamefont {Foster},
  \citenamefont {Kahn}, \citenamefont {Nguyen}, \citenamefont {Rodd},\ and\
  \citenamefont {Safdi}}]{Safdi_2021_Interferometer}%
  \BibitemOpen
  \bibfield  {author} {\bibinfo {author} {\bibfnamefont {J.~W.}\ \bibnamefont
  {Foster}}, \bibinfo {author} {\bibfnamefont {Y.}~\bibnamefont {Kahn}},
  \bibinfo {author} {\bibfnamefont {R.}~\bibnamefont {Nguyen}}, \bibinfo
  {author} {\bibfnamefont {N.~L.}\ \bibnamefont {Rodd}},\ and\ \bibinfo
  {author} {\bibfnamefont {B.~R.}\ \bibnamefont {Safdi}},\ }\bibfield  {title}
  {\bibinfo {title} {Dark matter interferometry},\ }\href
  {https://doi.org/10.1103/PhysRevD.103.076018} {\bibfield  {journal} {\bibinfo
   {journal} {Phys. Rev. D}\ }\textbf {\bibinfo {volume} {103}},\ \bibinfo
  {pages} {076018} (\bibinfo {year} {2021})}\BibitemShut {NoStop}%
\bibitem [{\citenamefont {Crescini}(2023)}]{crescini_fermionic_2023}%
  \BibitemOpen
  \bibfield  {author} {\bibinfo {author} {\bibfnamefont {N.}~\bibnamefont
  {Crescini}},\ }\bibfield  {title} {\bibinfo {title} {The {Fermionic} {Axion}
  {Interferometer}},\ }\href@noop {} {\bibfield  {journal} {\bibinfo  {journal}
  {arXiv preprint}\ } (\bibinfo {year} {2023})},\ \Eprint
  {https://arxiv.org/abs/2311.16364} {arXiv:2311.16364 [hep-ex]} \BibitemShut
  {NoStop}%
\bibitem [{\citenamefont {Padniuk}(2024)}]{Padniuk2024PhD}%
  \BibitemOpen
  \bibfield  {author} {\bibinfo {author} {\bibfnamefont {M.}~\bibnamefont
  {Padniuk}},\ }\emph {\bibinfo {title} {Wide-frequency-range atomic
  comagnetometry to search for spin-dependent interactions beyond the Standard
  Model}},\ \href@noop {} {Ph.D. thesis},\ \bibinfo  {school} {Jagiellonian
  University in Krak\'ow}, \bibinfo {address} {Poland} (\bibinfo {year}
  {2024})\BibitemShut {NoStop}%
\bibitem [{\citenamefont {Collaboration}\ \emph {et~al.}(2019)\citenamefont
  {Collaboration}, \citenamefont {Akiyama} \emph
  {et~al.}}]{collaboration_first_2019}%
  \BibitemOpen
  \bibfield  {author} {\bibinfo {author} {\bibfnamefont {T.~E. H.~T.}\
  \bibnamefont {Collaboration}}, \bibinfo {author} {\bibnamefont {Akiyama}},
  \emph {et~al.},\ }\bibfield  {title} {\bibinfo {title} {First {M87} {Event}
  {Horizon} {Telescope} {Results}. {I}. {The} {Shadow} of the {Supermassive}
  {Black} {Hole}},\ }\href {https://doi.org/10.3847/2041-8213/ab0ec7}
  {\bibfield  {journal} {\bibinfo  {journal} {Astrophys. J. Lett.}\ }\textbf
  {\bibinfo {volume} {875}},\ \bibinfo {pages} {L1} (\bibinfo {year}
  {2019})}\BibitemShut {NoStop}%
\bibitem [{\citenamefont {Budker}\ \emph {et~al.}(2023)\citenamefont {Budker},
  \citenamefont {Eby}, \citenamefont {Gorghetto}, \citenamefont {Jiang},\ and\
  \citenamefont {Perez}}]{Budker_2023_generic}%
  \BibitemOpen
  \bibfield  {author} {\bibinfo {author} {\bibfnamefont {D.}~\bibnamefont
  {Budker}}, \bibinfo {author} {\bibfnamefont {J.}~\bibnamefont {Eby}},
  \bibinfo {author} {\bibfnamefont {M.}~\bibnamefont {Gorghetto}}, \bibinfo
  {author} {\bibfnamefont {M.}~\bibnamefont {Jiang}},\ and\ \bibinfo {author}
  {\bibfnamefont {G.}~\bibnamefont {Perez}},\ }\bibfield  {title} {\bibinfo
  {title} {A generic formation mechanism of ultralight dark matter solar
  halos},\ }\href {https://doi.org/10.1088/1475-7516/2023/12/021} {\bibfield
  {journal} {\bibinfo  {journal} {J CAP}\ }\textbf {\bibinfo {volume} {2023}},\
  \bibinfo {pages} {021} (\bibinfo {year} {2023})}\BibitemShut {NoStop}%
\bibitem [{\citenamefont {O'Hare}\ \emph {et~al.}(2024)\citenamefont {O'Hare},
  \citenamefont {Pierobon},\ and\ \citenamefont {Redondo}}]{OHare_Axion_2024}%
  \BibitemOpen
  \bibfield  {author} {\bibinfo {author} {\bibfnamefont {C.~A.~J.}\
  \bibnamefont {O'Hare}}, \bibinfo {author} {\bibfnamefont {G.}~\bibnamefont
  {Pierobon}},\ and\ \bibinfo {author} {\bibfnamefont {J.}~\bibnamefont
  {Redondo}},\ }\bibfield  {title} {\bibinfo {title} {Axion minicluster streams
  in the solar neighborhood},\ }\href
  {https://doi.org/10.1103/PhysRevLett.133.081001} {\bibfield  {journal}
  {\bibinfo  {journal} {Phys. Rev. Lett.}\ }\textbf {\bibinfo {volume} {133}},\
  \bibinfo {pages} {081001} (\bibinfo {year} {2024})}\BibitemShut {NoStop}%
\bibitem [{\citenamefont {Kryemadhi}\ \emph {et~al.}(2023)\citenamefont
  {Kryemadhi}, \citenamefont {Maroudas}, \citenamefont {Mastronikolis},\ and\
  \citenamefont {Zioutas}}]{kryemadhi_gravitational_2023}%
  \BibitemOpen
  \bibfield  {author} {\bibinfo {author} {\bibfnamefont {A.}~\bibnamefont
  {Kryemadhi}}, \bibinfo {author} {\bibfnamefont {M.}~\bibnamefont {Maroudas}},
  \bibinfo {author} {\bibfnamefont {A.}~\bibnamefont {Mastronikolis}},\ and\
  \bibinfo {author} {\bibfnamefont {K.}~\bibnamefont {Zioutas}},\ }\bibfield
  {title} {\bibinfo {title} {Gravitational focusing effects on streaming dark
  matter as a new detection concept},\ }\href
  {https://doi.org/10.1103/PhysRevD.108.123043} {\bibfield  {journal} {\bibinfo
   {journal} {Phys. Rev. D}\ }\textbf {\bibinfo {volume} {108}},\ \bibinfo
  {pages} {123043} (\bibinfo {year} {2023})}\BibitemShut {NoStop}%
\bibitem [{\citenamefont {Khlopov}\ \emph {et~al.}(1999)\citenamefont
  {Khlopov}, \citenamefont {Sakharov},\ and\ \citenamefont
  {Sokoloff}}]{khlopov_nonlinear_1999}%
  \BibitemOpen
  \bibfield  {author} {\bibinfo {author} {\bibfnamefont {M.~Y.}\ \bibnamefont
  {Khlopov}}, \bibinfo {author} {\bibfnamefont {A.~S.}\ \bibnamefont
  {Sakharov}},\ and\ \bibinfo {author} {\bibfnamefont {D.~D.}\ \bibnamefont
  {Sokoloff}},\ }\bibfield  {title} {\bibinfo {title} {{The nonlinear
  modulation of the density distribution in standard axionic CDM and its
  cosmological impact}},\ }\href
  {https://doi.org/10.1016/S0920-5632(98)00511-8} {\bibfield  {journal}
  {\bibinfo  {journal} {Nucl. Phys. B Proc. Suppl.}\ }\textbf {\bibinfo
  {volume} {72}},\ \bibinfo {pages} {105} (\bibinfo {year} {1999})}\BibitemShut
  {NoStop}%
\bibitem [{\citenamefont {Baryakhtar}\ \emph {et~al.}(2021)\citenamefont
  {Baryakhtar}, \citenamefont {Galanis}, \citenamefont {Lasenby},\ and\
  \citenamefont {Simon}}]{Baryakhtar_Black_2021}%
  \BibitemOpen
  \bibfield  {author} {\bibinfo {author} {\bibfnamefont {M.}~\bibnamefont
  {Baryakhtar}}, \bibinfo {author} {\bibfnamefont {M.}~\bibnamefont {Galanis}},
  \bibinfo {author} {\bibfnamefont {R.}~\bibnamefont {Lasenby}},\ and\ \bibinfo
  {author} {\bibfnamefont {O.}~\bibnamefont {Simon}},\ }\bibfield  {title}
  {\bibinfo {title} {Black hole superradiance of self-interacting scalar
  fields},\ }\href {https://doi.org/10.1103/PhysRevD.103.095019} {\bibfield
  {journal} {\bibinfo  {journal} {Phys. Rev. D}\ }\textbf {\bibinfo {volume}
  {103}},\ \bibinfo {pages} {095019} (\bibinfo {year} {2021})}\BibitemShut
  {NoStop}%
\bibitem [{\citenamefont {Dailey}\ \emph {et~al.}(2020)\citenamefont {Dailey},
  \citenamefont {Bradley}, \citenamefont {Bradley}, \citenamefont {Kimball},
  \citenamefont {Sulai}, \citenamefont {Pustelny}, \citenamefont
  {Wickenbrock},\ and\ \citenamefont {Derevianko}}]{dailey_quantum_2020}%
  \BibitemOpen
  \bibfield  {author} {\bibinfo {author} {\bibfnamefont {C.}~\bibnamefont
  {Dailey}}, \bibinfo {author} {\bibfnamefont {C.}~\bibnamefont {Bradley}},
  \bibinfo {author} {\bibfnamefont {C.}~\bibnamefont {Bradley}}, \bibinfo
  {author} {\bibfnamefont {D.~F.~J.}\ \bibnamefont {Kimball}}, \bibinfo
  {author} {\bibfnamefont {I.~A.}\ \bibnamefont {Sulai}}, \bibinfo {author}
  {\bibfnamefont {S.}~\bibnamefont {Pustelny}}, \bibinfo {author}
  {\bibfnamefont {A.}~\bibnamefont {Wickenbrock}},\ and\ \bibinfo {author}
  {\bibfnamefont {A.}~\bibnamefont {Derevianko}},\ }\bibfield  {title}
  {\bibinfo {title} {Quantum sensor networks as exotic field telescopes for
  multi-messenger astronomy},\ }\href
  {https://doi.org/10.1038/s41550-020-01242-7} {\bibfield  {journal} {\bibinfo
  {journal} {Nat. Astron.}\ }\textbf {\bibinfo {volume} {5}},\ \bibinfo {pages}
  {150} (\bibinfo {year} {2020})}\BibitemShut {NoStop}%
\bibitem [{\citenamefont {Kornack}\ and\ \citenamefont
  {Romalis}(2002)}]{kornack_dynamics_2002}%
  \BibitemOpen
  \bibfield  {author} {\bibinfo {author} {\bibfnamefont {T.~W.}\ \bibnamefont
  {Kornack}}\ and\ \bibinfo {author} {\bibfnamefont {M.~V.}\ \bibnamefont
  {Romalis}},\ }\bibfield  {title} {\bibinfo {title} {Dynamics of {Two}
  {Overlapping} {Spin} {Ensembles} {Interacting} by {Spin} {Exchange}},\ }\href
  {https://doi.org/10.1103/PhysRevLett.89.253002} {\bibfield  {journal}
  {\bibinfo  {journal} {Phys. Rev. Lett.}\ }\textbf {\bibinfo {volume} {89}},\
  \bibinfo {pages} {253002} (\bibinfo {year} {2002})}\BibitemShut {NoStop}%
\bibitem [{\citenamefont {Wei}\ \emph {et~al.}(2023{\natexlab{a}})\citenamefont
  {Wei}, \citenamefont {Zhao}, \citenamefont {Fang}, \citenamefont {Xu},
  \citenamefont {Liu} \emph {et~al.}}]{wei_ultrasensitive_2023}%
  \BibitemOpen
  \bibfield  {author} {\bibinfo {author} {\bibfnamefont {K.}~\bibnamefont
  {Wei}}, \bibinfo {author} {\bibfnamefont {T.}~\bibnamefont {Zhao}}, \bibinfo
  {author} {\bibfnamefont {X.}~\bibnamefont {Fang}}, \bibinfo {author}
  {\bibfnamefont {Z.}~\bibnamefont {Xu}}, \bibinfo {author} {\bibfnamefont
  {C.}~\bibnamefont {Liu}}, \emph {et~al.},\ }\bibfield  {title} {\bibinfo
  {title} {Ultrasensitive {Atomic} {Comagnetometer} with {Enhanced} {Nuclear}
  {Spin} {Coherence}},\ }\href {https://doi.org/10.1103/PhysRevLett.130.063201}
  {\bibfield  {journal} {\bibinfo  {journal} {Phys. Rev. Lett.}\ }\textbf
  {\bibinfo {volume} {130}},\ \bibinfo {pages} {063201} (\bibinfo {year}
  {2023}{\natexlab{a}})}\BibitemShut {NoStop}%
\bibitem [{\citenamefont {Klinger}\ \emph {et~al.}(2023)\citenamefont
  {Klinger}, \citenamefont {Liu}, \citenamefont {Padniuk}, \citenamefont
  {Engler}, \citenamefont {Kornack} \emph {et~al.}}]{klingerPRA2023}%
  \BibitemOpen
  \bibfield  {author} {\bibinfo {author} {\bibfnamefont {E.}~\bibnamefont
  {Klinger}}, \bibinfo {author} {\bibfnamefont {T.}~\bibnamefont {Liu}},
  \bibinfo {author} {\bibfnamefont {M.}~\bibnamefont {Padniuk}}, \bibinfo
  {author} {\bibfnamefont {M.}~\bibnamefont {Engler}}, \bibinfo {author}
  {\bibfnamefont {T.}~\bibnamefont {Kornack}}, \emph {et~al.},\ }\bibfield
  {title} {\bibinfo {title} {Optimization of nuclear polarization in an
  alkali-noble gas comagnetometer},\ }\href
  {https://doi.org/10.1103/PhysRevApplied.19.044092} {\bibfield  {journal}
  {\bibinfo  {journal} {Phys. Rev. Appl.}\ }\textbf {\bibinfo {volume} {19}},\
  \bibinfo {pages} {044092} (\bibinfo {year} {2023})}\BibitemShut {NoStop}%
\bibitem [{\citenamefont {Padniuk}\ \emph {et~al.}(2022)\citenamefont
  {Padniuk}, \citenamefont {Kopciuch}, \citenamefont {Cipolletti},
  \citenamefont {Wickenbrock}, \citenamefont {Budker},\ and\ \citenamefont
  {Pustelny}}]{padniuk_response_2022}%
  \BibitemOpen
  \bibfield  {author} {\bibinfo {author} {\bibfnamefont {M.}~\bibnamefont
  {Padniuk}}, \bibinfo {author} {\bibfnamefont {M.}~\bibnamefont {Kopciuch}},
  \bibinfo {author} {\bibfnamefont {R.}~\bibnamefont {Cipolletti}}, \bibinfo
  {author} {\bibfnamefont {A.}~\bibnamefont {Wickenbrock}}, \bibinfo {author}
  {\bibfnamefont {D.}~\bibnamefont {Budker}},\ and\ \bibinfo {author}
  {\bibfnamefont {S.}~\bibnamefont {Pustelny}},\ }\bibfield  {title} {\bibinfo
  {title} {Response of atomic spin-based sensors to magnetic and nonmagnetic
  perturbations},\ }\href {https://doi.org/10.1038/s41598-021-03609-w}
  {\bibfield  {journal} {\bibinfo  {journal} {Sci. Rep.}\ }\textbf {\bibinfo
  {volume} {12}},\ \bibinfo {pages} {324} (\bibinfo {year} {2022})}\BibitemShut
  {NoStop}%
\bibitem [{\citenamefont {Padniuk}\ \emph {et~al.}(2024)\citenamefont
  {Padniuk}, \citenamefont {Klinger}, \citenamefont {Łukasiewicz},
  \citenamefont {Gavilan-Martin}, \citenamefont {Liu} \emph
  {et~al.}}]{padniuk_universal_2024}%
  \BibitemOpen
  \bibfield  {author} {\bibinfo {author} {\bibfnamefont {M.}~\bibnamefont
  {Padniuk}}, \bibinfo {author} {\bibfnamefont {E.}~\bibnamefont {Klinger}},
  \bibinfo {author} {\bibfnamefont {G.}~\bibnamefont {Łukasiewicz}}, \bibinfo
  {author} {\bibfnamefont {D.}~\bibnamefont {Gavilan-Martin}}, \bibinfo
  {author} {\bibfnamefont {T.}~\bibnamefont {Liu}}, \emph {et~al.},\ }\bibfield
   {title} {\bibinfo {title} {Universal determination of comagnetometer
  response to spin couplings},\ }\href
  {https://doi.org/10.1103/PhysRevResearch.6.013339} {\bibfield  {journal}
  {\bibinfo  {journal} {Phys. Rev. Res.}\ }\textbf {\bibinfo {volume} {6}},\
  \bibinfo {pages} {013339} (\bibinfo {year} {2024})}\BibitemShut {NoStop}%
\bibitem [{\citenamefont {Lee}\ \emph {et~al.}(2023)\citenamefont {Lee},
  \citenamefont {Lisanti}, \citenamefont {Terrano},\ and\ \citenamefont
  {Romalis}}]{lee_laboratory_2023}%
  \BibitemOpen
  \bibfield  {author} {\bibinfo {author} {\bibfnamefont {J.}~\bibnamefont
  {Lee}}, \bibinfo {author} {\bibfnamefont {M.}~\bibnamefont {Lisanti}},
  \bibinfo {author} {\bibfnamefont {W.~A.}\ \bibnamefont {Terrano}},\ and\
  \bibinfo {author} {\bibfnamefont {M.}~\bibnamefont {Romalis}},\ }\bibfield
  {title} {\bibinfo {title} {Laboratory {Constraints} on the {Neutron}-{Spin}
  {Coupling} of {feV}-{Scale} {Axions}},\ }\href
  {https://doi.org/10.1103/PhysRevX.13.011050} {\bibfield  {journal} {\bibinfo
  {journal} {Phys. Rev. X}\ }\textbf {\bibinfo {volume} {13}},\ \bibinfo
  {pages} {011050} (\bibinfo {year} {2023})}\BibitemShut {NoStop}%
\bibitem [{\citenamefont {Wei}\ \emph {et~al.}(2023{\natexlab{b}})\citenamefont
  {Wei}, \citenamefont {Xu}, \citenamefont {He}, \citenamefont {Ma},
  \citenamefont {Heng} \emph {et~al.}}]{wei_dark_2023}%
  \BibitemOpen
  \bibfield  {author} {\bibinfo {author} {\bibfnamefont {K.}~\bibnamefont
  {Wei}}, \bibinfo {author} {\bibfnamefont {Z.}~\bibnamefont {Xu}}, \bibinfo
  {author} {\bibfnamefont {Y.}~\bibnamefont {He}}, \bibinfo {author}
  {\bibfnamefont {X.}~\bibnamefont {Ma}}, \bibinfo {author} {\bibfnamefont
  {X.}~\bibnamefont {Heng}}, \emph {et~al.},\ }\bibfield  {title} {\bibinfo
  {title} {Dark matter search with a strongly-coupled hybrid spin system},\
  }\href@noop {} {\bibfield  {journal} {\bibinfo  {journal} {arXiv preprint}\ }
  (\bibinfo {year} {2023}{\natexlab{b}})},\ \Eprint
  {https://arxiv.org/abs/2306.08039} {arXiv:2306.08039 [hep-ex]} \BibitemShut
  {NoStop}%
\bibitem [{\citenamefont {Xu}\ \emph {et~al.}(2024)\citenamefont {Xu},
  \citenamefont {Ma}, \citenamefont {Wei}, \citenamefont {He}, \citenamefont
  {Heng} \emph {et~al.}}]{xu_constraining_2024}%
  \BibitemOpen
  \bibfield  {author} {\bibinfo {author} {\bibfnamefont {Z.}~\bibnamefont
  {Xu}}, \bibinfo {author} {\bibfnamefont {X.}~\bibnamefont {Ma}}, \bibinfo
  {author} {\bibfnamefont {K.}~\bibnamefont {Wei}}, \bibinfo {author}
  {\bibfnamefont {Y.}~\bibnamefont {He}}, \bibinfo {author} {\bibfnamefont
  {X.}~\bibnamefont {Heng}}, \emph {et~al.},\ }\bibfield  {title} {\bibinfo
  {title} {Constraining ultralight dark matter through an accelerated resonant
  search},\ }\href {https://doi.org/10.1038/s42005-024-01713-7} {\bibfield
  {journal} {\bibinfo  {journal} {Commun. Phys.}\ }\textbf {\bibinfo {volume}
  {7}},\ \bibinfo {pages} {1} (\bibinfo {year} {2024})}\BibitemShut {NoStop}%
\bibitem [{\citenamefont {Workman}\ \emph {et~al.}(2022)\citenamefont {Workman}
  \emph {et~al.}}]{Workman_2022_PDG}%
  \BibitemOpen
  \bibfield  {author} {\bibinfo {author} {\bibfnamefont {R.~L.}\ \bibnamefont
  {Workman}} \emph {et~al.} (\bibinfo {collaboration} {Particle Data Group}),\
  }\bibfield  {title} {\bibinfo {title} {{Review of Particle Physics}},\ }\href
  {https://doi.org/10.1093/ptep/ptac097} {\bibfield  {journal} {\bibinfo
  {journal} {PTEP}\ }\textbf {\bibinfo {volume} {2022}},\ \bibinfo {pages}
  {083C01} (\bibinfo {year} {2022})}\BibitemShut {NoStop}%
\bibitem [{\citenamefont {Chadha-Day}\ \emph {et~al.}(2022)\citenamefont
  {Chadha-Day}, \citenamefont {Ellis},\ and\ \citenamefont
  {Marsh}}]{chadha-day_axion_2022}%
  \BibitemOpen
  \bibfield  {author} {\bibinfo {author} {\bibfnamefont {F.}~\bibnamefont
  {Chadha-Day}}, \bibinfo {author} {\bibfnamefont {J.}~\bibnamefont {Ellis}},\
  and\ \bibinfo {author} {\bibfnamefont {D.~J.~E.}\ \bibnamefont {Marsh}},\
  }\bibfield  {title} {\bibinfo {title} {Axion dark matter: {What} is it and
  why now?},\ }\href {https://doi.org/10.1126/sciadv.abj3618} {\bibfield
  {journal} {\bibinfo  {journal} {Sci. Adv.}\ }\textbf {\bibinfo {volume}
  {8}},\ \bibinfo {pages} {eabj3618} (\bibinfo {year} {2022})}\BibitemShut
  {NoStop}%
\bibitem [{\citenamefont {Foster}\ \emph {et~al.}(2018)\citenamefont {Foster},
  \citenamefont {Rodd},\ and\ \citenamefont {Safdi}}]{foster_revealing_2018}%
  \BibitemOpen
  \bibfield  {author} {\bibinfo {author} {\bibfnamefont {J.~W.}\ \bibnamefont
  {Foster}}, \bibinfo {author} {\bibfnamefont {N.~L.}\ \bibnamefont {Rodd}},\
  and\ \bibinfo {author} {\bibfnamefont {B.~R.}\ \bibnamefont {Safdi}},\
  }\bibfield  {title} {\bibinfo {title} {Revealing the dark matter halo with
  axion direct detection},\ }\href {https://doi.org/10.1103/PhysRevD.97.123006}
  {\bibfield  {journal} {\bibinfo  {journal} {Phys. Rev. D}\ }\textbf {\bibinfo
  {volume} {97}},\ \bibinfo {pages} {123006} (\bibinfo {year}
  {2018})}\BibitemShut {NoStop}%
\bibitem [{\citenamefont {Centers}\ \emph {et~al.}(2021)\citenamefont
  {Centers}, \citenamefont {Blanchard}, \citenamefont {Conrad}, \citenamefont
  {Figueroa}, \citenamefont {Garcon} \emph {et~al.}}]{centers_stochastic_2021}%
  \BibitemOpen
  \bibfield  {author} {\bibinfo {author} {\bibfnamefont {G.~P.}\ \bibnamefont
  {Centers}}, \bibinfo {author} {\bibfnamefont {J.~W.}\ \bibnamefont
  {Blanchard}}, \bibinfo {author} {\bibfnamefont {J.}~\bibnamefont {Conrad}},
  \bibinfo {author} {\bibfnamefont {N.~L.}\ \bibnamefont {Figueroa}}, \bibinfo
  {author} {\bibfnamefont {A.}~\bibnamefont {Garcon}}, \emph {et~al.},\
  }\bibfield  {title} {\bibinfo {title} {Stochastic fluctuations of bosonic
  dark matter},\ }\href {https://doi.org/10.1038/s41467-021-27632-7} {\bibfield
   {journal} {\bibinfo  {journal} {Nat. Commun.}\ }\textbf {\bibinfo {volume}
  {12}},\ \bibinfo {pages} {7321} (\bibinfo {year} {2021})}\BibitemShut
  {NoStop}%
\bibitem [{\citenamefont {Lisanti}\ \emph {et~al.}(2021)\citenamefont
  {Lisanti}, \citenamefont {Moschella},\ and\ \citenamefont
  {Terrano}}]{lisanti_stochastic_2021}%
  \BibitemOpen
  \bibfield  {author} {\bibinfo {author} {\bibfnamefont {M.}~\bibnamefont
  {Lisanti}}, \bibinfo {author} {\bibfnamefont {M.}~\bibnamefont {Moschella}},\
  and\ \bibinfo {author} {\bibfnamefont {W.}~\bibnamefont {Terrano}},\
  }\bibfield  {title} {\bibinfo {title} {Stochastic properties of ultralight
  scalar field gradients},\ }\href
  {https://doi.org/10.1103/PhysRevD.104.055037} {\bibfield  {journal} {\bibinfo
   {journal} {Phys. Rev. D}\ }\textbf {\bibinfo {volume} {104}},\ \bibinfo
  {pages} {055037} (\bibinfo {year} {2021})}\BibitemShut {NoStop}%
\bibitem [{\citenamefont {Gramolin}\ \emph {et~al.}(2022)\citenamefont
  {Gramolin}, \citenamefont {Wickenbrock}, \citenamefont {Aybas}, \citenamefont
  {Bekker}, \citenamefont {Budker} \emph {et~al.}}]{gramolin_spectral_2022}%
  \BibitemOpen
  \bibfield  {author} {\bibinfo {author} {\bibfnamefont {A.~V.}\ \bibnamefont
  {Gramolin}}, \bibinfo {author} {\bibfnamefont {A.}~\bibnamefont
  {Wickenbrock}}, \bibinfo {author} {\bibfnamefont {D.}~\bibnamefont {Aybas}},
  \bibinfo {author} {\bibfnamefont {H.}~\bibnamefont {Bekker}}, \bibinfo
  {author} {\bibfnamefont {D.}~\bibnamefont {Budker}}, \emph {et~al.},\
  }\bibfield  {title} {\bibinfo {title} {Spectral signatures of axionlike dark
  matter},\ }\href {https://doi.org/10.1103/PhysRevD.105.035029} {\bibfield
  {journal} {\bibinfo  {journal} {Phys. Rev. D}\ }\textbf {\bibinfo {volume}
  {105}},\ \bibinfo {pages} {035029} (\bibinfo {year} {2022})}\BibitemShut
  {NoStop}%
\bibitem [{\citenamefont {Flambaum}\ and\ \citenamefont
  {Samsonov}(2023)}]{flambaum_fluctuations_2023}%
  \BibitemOpen
  \bibfield  {author} {\bibinfo {author} {\bibfnamefont {V.}~\bibnamefont
  {Flambaum}}\ and\ \bibinfo {author} {\bibfnamefont {I.}~\bibnamefont
  {Samsonov}},\ }\bibfield  {title} {\bibinfo {title} {Fluctuations of atomic
  energy levels due to axion dark matter},\ }\href
  {https://doi.org/10.1103/PhysRevD.108.075022} {\bibfield  {journal} {\bibinfo
   {journal} {Phys. Rev. D}\ }\textbf {\bibinfo {volume} {108}},\ \bibinfo
  {pages} {075022} (\bibinfo {year} {2023})}\BibitemShut {NoStop}%
\bibitem [{\citenamefont {Kimball}(2015)}]{kimball_nuclear_2015}%
  \BibitemOpen
  \bibfield  {author} {\bibinfo {author} {\bibfnamefont {D.~F.~J.}\
  \bibnamefont {Kimball}},\ }\bibfield  {title} {\bibinfo {title} {Nuclear spin
  content and constraints on exotic spin-dependent couplings},\ }\href
  {https://doi.org/10.1088/1367-2630/17/7/073008} {\bibfield  {journal}
  {\bibinfo  {journal} {New J. Phys.}\ }\textbf {\bibinfo {volume} {17}},\
  \bibinfo {pages} {073008} (\bibinfo {year} {2015})},\ \bibinfo {note}
  {publisher: IOP Publishing}\BibitemShut {NoStop}%
\bibitem [{\citenamefont {Derevianko}(2018)}]{derevianko_detecting_2018}%
  \BibitemOpen
  \bibfield  {author} {\bibinfo {author} {\bibfnamefont {A.}~\bibnamefont
  {Derevianko}},\ }\bibfield  {title} {\bibinfo {title} {Detecting dark-matter
  waves with a network of precision-measurement tools},\ }\href
  {https://doi.org/10.1103/PhysRevA.97.042506} {\bibfield  {journal} {\bibinfo
  {journal} {Phys. Rev. A}\ }\textbf {\bibinfo {volume} {97}},\ \bibinfo
  {pages} {042506} (\bibinfo {year} {2018})}\BibitemShut {NoStop}%
\bibitem [{\citenamefont {Jackson~Kimball}\ \emph {et~al.}(2016)\citenamefont
  {Jackson~Kimball}, \citenamefont {Dudley}, \citenamefont {Li}, \citenamefont
  {Thulasi}, \citenamefont {Pustelny}, \citenamefont {Budker},\ and\
  \citenamefont {Zolotorev}}]{Kimball_shielding_2016}%
  \BibitemOpen
  \bibfield  {author} {\bibinfo {author} {\bibfnamefont {D.~F.}\ \bibnamefont
  {Jackson~Kimball}}, \bibinfo {author} {\bibfnamefont {J.}~\bibnamefont
  {Dudley}}, \bibinfo {author} {\bibfnamefont {Y.}~\bibnamefont {Li}}, \bibinfo
  {author} {\bibfnamefont {S.}~\bibnamefont {Thulasi}}, \bibinfo {author}
  {\bibfnamefont {S.}~\bibnamefont {Pustelny}}, \bibinfo {author}
  {\bibfnamefont {D.}~\bibnamefont {Budker}},\ and\ \bibinfo {author}
  {\bibfnamefont {M.}~\bibnamefont {Zolotorev}},\ }\bibfield  {title} {\bibinfo
  {title} {Magnetic shielding and exotic spin-dependent interactions},\ }\href
  {https://doi.org/10.1103/PhysRevD.94.082005} {\bibfield  {journal} {\bibinfo
  {journal} {Phys. Rev. D}\ }\textbf {\bibinfo {volume} {94}},\ \bibinfo
  {pages} {082005} (\bibinfo {year} {2016})}\BibitemShut {NoStop}%
\bibitem [{\citenamefont {Read}(2002)}]{read_presentation_2002}%
  \BibitemOpen
  \bibfield  {author} {\bibinfo {author} {\bibfnamefont {A.~L.}\ \bibnamefont
  {Read}},\ }\bibfield  {title} {\bibinfo {title} {Presentation of search
  results: the {CLs} technique},\ }\href
  {https://doi.org/10.1088/0954-3899/28/10/313} {\bibfield  {journal} {\bibinfo
   {journal} {J. Phys. G: Nucl. Part. Phys.}\ }\textbf {\bibinfo {volume}
  {28}},\ \bibinfo {pages} {2693} (\bibinfo {year} {2002})}\BibitemShut
  {NoStop}%
\bibitem [{\citenamefont {Bloch}\ \emph {et~al.}(2020)\citenamefont {Bloch},
  \citenamefont {Hochberg}, \citenamefont {Kuflik},\ and\ \citenamefont
  {Volansky}}]{bloch_axion-like_2020}%
  \BibitemOpen
  \bibfield  {author} {\bibinfo {author} {\bibfnamefont {I.~M.}\ \bibnamefont
  {Bloch}}, \bibinfo {author} {\bibfnamefont {Y.}~\bibnamefont {Hochberg}},
  \bibinfo {author} {\bibfnamefont {E.}~\bibnamefont {Kuflik}},\ and\ \bibinfo
  {author} {\bibfnamefont {T.}~\bibnamefont {Volansky}},\ }\bibfield  {title}
  {\bibinfo {title} {Axion-like relics: new constraints from old comagnetometer
  data},\ }\href {https://doi.org/10.1007/JHEP01(2020)167} {\bibfield
  {journal} {\bibinfo  {journal} {J HEP}\ }\textbf {\bibinfo {volume} {2020}},\
  \bibinfo {pages} {167} (\bibinfo {year} {2020})}\BibitemShut {NoStop}%
\bibitem [{\citenamefont {Vasilakis}\ \emph {et~al.}(2009)\citenamefont
  {Vasilakis}, \citenamefont {Brown}, \citenamefont {Kornack},\ and\
  \citenamefont {Romalis}}]{vasilakis_limits_2009}%
  \BibitemOpen
  \bibfield  {author} {\bibinfo {author} {\bibfnamefont {G.}~\bibnamefont
  {Vasilakis}}, \bibinfo {author} {\bibfnamefont {J.~M.}\ \bibnamefont
  {Brown}}, \bibinfo {author} {\bibfnamefont {T.~W.}\ \bibnamefont {Kornack}},\
  and\ \bibinfo {author} {\bibfnamefont {M.~V.}\ \bibnamefont {Romalis}},\
  }\bibfield  {title} {\bibinfo {title} {Limits on {New} {Long} {Range}
  {Nuclear} {Spin}-{Dependent} {Forces} {Set} with a {K} − $^3${He}
  {Comagnetometer}},\ }\href {https://doi.org/10.1103/PhysRevLett.103.261801}
  {\bibfield  {journal} {\bibinfo  {journal} {Phys. Rev. Lett.}\ }\textbf
  {\bibinfo {volume} {103}},\ \bibinfo {pages} {261801} (\bibinfo {year}
  {2009})}\BibitemShut {NoStop}%
\bibitem [{\citenamefont {Brown}\ \emph {et~al.}(2010)\citenamefont {Brown},
  \citenamefont {Smullin}, \citenamefont {Kornack},\ and\ \citenamefont
  {Romalis}}]{brown_new_2010}%
  \BibitemOpen
  \bibfield  {author} {\bibinfo {author} {\bibfnamefont {J.~M.}\ \bibnamefont
  {Brown}}, \bibinfo {author} {\bibfnamefont {S.~J.}\ \bibnamefont {Smullin}},
  \bibinfo {author} {\bibfnamefont {T.~W.}\ \bibnamefont {Kornack}},\ and\
  \bibinfo {author} {\bibfnamefont {M.~V.}\ \bibnamefont {Romalis}},\
  }\bibfield  {title} {\bibinfo {title} {New {Limit} on {Lorentz}- and
  {CPT}-{Violating} {Neutron} {Spin} {Interactions}},\ }\href
  {https://doi.org/10.1103/physrevlett.105.151604} {\bibfield  {journal}
  {\bibinfo  {journal} {Phys. Rev. Lett.}\ }\textbf {\bibinfo {volume} {105}},\
  \bibinfo {pages} {151604} (\bibinfo {year} {2010})}\BibitemShut {NoStop}%
\bibitem [{\citenamefont {Lee}\ \emph {et~al.}(2018)\citenamefont {Lee},
  \citenamefont {Almasi},\ and\ \citenamefont {Romalis}}]{lee_improved_2018}%
  \BibitemOpen
  \bibfield  {author} {\bibinfo {author} {\bibfnamefont {J.}~\bibnamefont
  {Lee}}, \bibinfo {author} {\bibfnamefont {A.}~\bibnamefont {Almasi}},\ and\
  \bibinfo {author} {\bibfnamefont {M.}~\bibnamefont {Romalis}},\ }\bibfield
  {title} {\bibinfo {title} {Improved {Limits} on {Spin}-{Mass}
  {Interactions}},\ }\href {https://doi.org/10.1103/PhysRevLett.120.161801}
  {\bibfield  {journal} {\bibinfo  {journal} {Phys. Rev. Lett.}\ }\textbf
  {\bibinfo {volume} {120}},\ \bibinfo {pages} {161801} (\bibinfo {year}
  {2018})}\BibitemShut {NoStop}%
\bibitem [{\citenamefont {Afach}\ \emph {et~al.}(2023)\citenamefont {Afach},
  \citenamefont {Aybas~Tumturk}, \citenamefont {Bekker}, \citenamefont
  {Buchler}, \citenamefont {Budker} \emph {et~al.}}]{afach_what_nodate}%
  \BibitemOpen
  \bibfield  {author} {\bibinfo {author} {\bibfnamefont {S.}~\bibnamefont
  {Afach}}, \bibinfo {author} {\bibfnamefont {D.}~\bibnamefont
  {Aybas~Tumturk}}, \bibinfo {author} {\bibfnamefont {H.}~\bibnamefont
  {Bekker}}, \bibinfo {author} {\bibfnamefont {B.~C.}\ \bibnamefont {Buchler}},
  \bibinfo {author} {\bibfnamefont {D.}~\bibnamefont {Budker}}, \emph
  {et~al.},\ }\bibfield  {title} {\bibinfo {title} {What {Can} a {GNOME} {Do}?
  {Search} {Targets} for the {Global} {Network} of {Optical} {Magnetometers}
  for {Exotic} {Physics} {Searches}},\ }\href
  {https://doi.org/10.1002/andp.202300083} {\bibfield  {journal} {\bibinfo
  {journal} {Ann. Phys.}\ }\textbf {\bibinfo {volume} {536}},\ \bibinfo {pages}
  {2300083} (\bibinfo {year} {2023})}\BibitemShut {NoStop}%
\bibitem [{\citenamefont {Garcon}\ \emph {et~al.}(2019)\citenamefont {Garcon},
  \citenamefont {Blanchard}, \citenamefont {Centers}, \citenamefont {Figueroa},
  \citenamefont {Graham} \emph {et~al.}}]{garcon_constraints_2019}%
  \BibitemOpen
  \bibfield  {author} {\bibinfo {author} {\bibfnamefont {A.}~\bibnamefont
  {Garcon}}, \bibinfo {author} {\bibfnamefont {J.~W.}\ \bibnamefont
  {Blanchard}}, \bibinfo {author} {\bibfnamefont {G.~P.}\ \bibnamefont
  {Centers}}, \bibinfo {author} {\bibfnamefont {N.~L.}\ \bibnamefont
  {Figueroa}}, \bibinfo {author} {\bibfnamefont {P.~W.}\ \bibnamefont
  {Graham}}, \emph {et~al.},\ }\bibfield  {title} {\bibinfo {title}
  {Constraints on bosonic dark matter from ultralow-field nuclear magnetic
  resonance},\ }\href {https://doi.org/10.1126/sciadv.aax4539} {\bibfield
  {journal} {\bibinfo  {journal} {Sci. Adv.}\ }\textbf {\bibinfo {volume}
  {5}},\ \bibinfo {pages} {eaax4539} (\bibinfo {year} {2019})}\BibitemShut
  {NoStop}%
\bibitem [{\citenamefont {Wu}\ \emph {et~al.}(2019)\citenamefont {Wu},
  \citenamefont {Blanchard}, \citenamefont {Centers}, \citenamefont {Figueroa},
  \citenamefont {Garcon} \emph {et~al.}}]{wu_search_2019}%
  \BibitemOpen
  \bibfield  {author} {\bibinfo {author} {\bibfnamefont {T.}~\bibnamefont
  {Wu}}, \bibinfo {author} {\bibfnamefont {J.~W.}\ \bibnamefont {Blanchard}},
  \bibinfo {author} {\bibfnamefont {G.~P.}\ \bibnamefont {Centers}}, \bibinfo
  {author} {\bibfnamefont {N.~L.}\ \bibnamefont {Figueroa}}, \bibinfo {author}
  {\bibfnamefont {A.}~\bibnamefont {Garcon}}, \emph {et~al.},\ }\bibfield
  {title} {\bibinfo {title} {Search for {Axionlike} {Dark} {Matter} with a
  {Liquid}-{State} {Nuclear} {Spin} {Comagnetometer}},\ }\href
  {https://doi.org/10.1103/PhysRevLett.122.191302} {\bibfield  {journal}
  {\bibinfo  {journal} {Phys. Rev. Lett.}\ }\textbf {\bibinfo {volume} {122}},\
  \bibinfo {pages} {191302} (\bibinfo {year} {2019})}\BibitemShut {NoStop}%
\bibitem [{\citenamefont {O'Hare}(2020)}]{AxionLimits}%
  \BibitemOpen
  \bibfield  {author} {\bibinfo {author} {\bibfnamefont {C.}~\bibnamefont
  {O'Hare}},\ }\href {https://doi.org/10.5281/zenodo.3932430} {\bibinfo {title}
  {cajohare/axionlimits: Axionlimits}},\ \bibinfo {howpublished}
  {\url{https://cajohare.github.io/AxionLimits/}} (\bibinfo {year}
  {2020})\BibitemShut {NoStop}%
\bibitem [{\citenamefont {Abel}\ \emph {et~al.}(2017)\citenamefont {Abel},
  \citenamefont {Ayres}, \citenamefont {Ban}, \citenamefont {Bison},
  \citenamefont {Bodek} \emph {et~al.}}]{abel_search_2017}%
  \BibitemOpen
  \bibfield  {author} {\bibinfo {author} {\bibfnamefont {C.}~\bibnamefont
  {Abel}}, \bibinfo {author} {\bibfnamefont {N.}~\bibnamefont {Ayres}},
  \bibinfo {author} {\bibfnamefont {G.}~\bibnamefont {Ban}}, \bibinfo {author}
  {\bibfnamefont {G.}~\bibnamefont {Bison}}, \bibinfo {author} {\bibfnamefont
  {K.}~\bibnamefont {Bodek}}, \emph {et~al.},\ }\bibfield  {title} {\bibinfo
  {title} {Search for {Axionlike} {Dark} {Matter} through {Nuclear} {Spin}
  {Precession} in {Electric} and {Magnetic} {Fields}},\ }\href
  {https://doi.org/10.1103/PhysRevX.7.041034} {\bibfield  {journal} {\bibinfo
  {journal} {Phys. Rev. X}\ }\textbf {\bibinfo {volume} {7}},\ \bibinfo {pages}
  {041034} (\bibinfo {year} {2017})}\BibitemShut {NoStop}%
\bibitem [{\citenamefont {Abel}\ \emph {et~al.}(2023)\citenamefont {Abel},
  \citenamefont {Ayres}, \citenamefont {Ban}, \citenamefont {Bison},
  \citenamefont {Bodek} \emph {et~al.}}]{abel_search_2023}%
  \BibitemOpen
  \bibfield  {author} {\bibinfo {author} {\bibfnamefont {C.}~\bibnamefont
  {Abel}}, \bibinfo {author} {\bibfnamefont {N.~J.}\ \bibnamefont {Ayres}},
  \bibinfo {author} {\bibfnamefont {G.}~\bibnamefont {Ban}}, \bibinfo {author}
  {\bibfnamefont {G.}~\bibnamefont {Bison}}, \bibinfo {author} {\bibfnamefont
  {K.}~\bibnamefont {Bodek}}, \emph {et~al.},\ }\bibfield  {title} {\bibinfo
  {title} {Search for ultralight axion dark matter in a side-band analysis of a
  $^{199}${Hg} free-spin precession signal},\ }\href
  {https://doi.org/10.21468/SciPostPhys.15.2.058} {\bibfield  {journal}
  {\bibinfo  {journal} {SciPost Phys.}\ }\textbf {\bibinfo {volume} {15}},\
  \bibinfo {pages} {058} (\bibinfo {year} {2023})}\BibitemShut {NoStop}%
\bibitem [{\citenamefont {Bhusal}\ \emph {et~al.}(2021)\citenamefont {Bhusal},
  \citenamefont {Houston},\ and\ \citenamefont {Li}}]{bhusal_searching_2021}%
  \BibitemOpen
  \bibfield  {author} {\bibinfo {author} {\bibfnamefont {A.}~\bibnamefont
  {Bhusal}}, \bibinfo {author} {\bibfnamefont {N.}~\bibnamefont {Houston}},\
  and\ \bibinfo {author} {\bibfnamefont {T.}~\bibnamefont {Li}},\ }\bibfield
  {title} {\bibinfo {title} {Searching for {Solar} {Axions} {Using} {Data} from
  the {Sudbury} {Neutrino} {Observatory}},\ }\href
  {https://doi.org/10.1103/PhysRevLett.126.091601} {\bibfield  {journal}
  {\bibinfo  {journal} {Phys. Rev. Lett.}\ }\textbf {\bibinfo {volume} {126}},\
  \bibinfo {pages} {091601} (\bibinfo {year} {2021})}\BibitemShut {NoStop}%
\bibitem [{\citenamefont {Bloch}\ \emph {et~al.}(2023)\citenamefont {Bloch},
  \citenamefont {Shaham}, \citenamefont {Hochberg}, \citenamefont {Kuflik},
  \citenamefont {Volansky},\ and\ \citenamefont {Katz}}]{bloch_nasduck_2022}%
  \BibitemOpen
  \bibfield  {author} {\bibinfo {author} {\bibfnamefont {I.~M.}\ \bibnamefont
  {Bloch}}, \bibinfo {author} {\bibfnamefont {R.}~\bibnamefont {Shaham}},
  \bibinfo {author} {\bibfnamefont {Y.}~\bibnamefont {Hochberg}}, \bibinfo
  {author} {\bibfnamefont {E.}~\bibnamefont {Kuflik}}, \bibinfo {author}
  {\bibfnamefont {T.}~\bibnamefont {Volansky}},\ and\ \bibinfo {author}
  {\bibfnamefont {O.}~\bibnamefont {Katz}},\ }\bibfield  {title} {\bibinfo
  {title} {Constraints on axion-like dark matter from a serf comagnetometer},\
  }\href {https://doi.org/10.1038/s41467-023-41162-4} {\bibfield  {journal}
  {\bibinfo  {journal} {Nat. Commun.}\ }\textbf {\bibinfo {volume} {14}},\
  \bibinfo {pages} {5784} (\bibinfo {year} {2023})}\BibitemShut {NoStop}%
\bibitem [{\citenamefont {Jiang}\ \emph {et~al.}(2021)\citenamefont {Jiang},
  \citenamefont {Su}, \citenamefont {Garcon}, \citenamefont {Peng},\ and\
  \citenamefont {Budker}}]{jiang_search_2021}%
  \BibitemOpen
  \bibfield  {author} {\bibinfo {author} {\bibfnamefont {M.}~\bibnamefont
  {Jiang}}, \bibinfo {author} {\bibfnamefont {H.}~\bibnamefont {Su}}, \bibinfo
  {author} {\bibfnamefont {A.}~\bibnamefont {Garcon}}, \bibinfo {author}
  {\bibfnamefont {X.}~\bibnamefont {Peng}},\ and\ \bibinfo {author}
  {\bibfnamefont {D.}~\bibnamefont {Budker}},\ }\bibfield  {title} {\bibinfo
  {title} {Search for axion-like dark matter with spin-based amplifiers},\
  }\href {https://doi.org/10.1038/s41567-021-01392-z} {\bibfield  {journal}
  {\bibinfo  {journal} {Nat. Phys.}\ }\textbf {\bibinfo {volume} {17}},\
  \bibinfo {pages} {1402} (\bibinfo {year} {2021})}\BibitemShut {NoStop}%
\bibitem [{\citenamefont {Buschmann}\ \emph {et~al.}(2022)\citenamefont
  {Buschmann}, \citenamefont {Dessert}, \citenamefont {Foster}, \citenamefont
  {Long},\ and\ \citenamefont {Safdi}}]{buschmann_upper_2022}%
  \BibitemOpen
  \bibfield  {author} {\bibinfo {author} {\bibfnamefont {M.}~\bibnamefont
  {Buschmann}}, \bibinfo {author} {\bibfnamefont {C.}~\bibnamefont {Dessert}},
  \bibinfo {author} {\bibfnamefont {J.~W.}\ \bibnamefont {Foster}}, \bibinfo
  {author} {\bibfnamefont {A.~J.}\ \bibnamefont {Long}},\ and\ \bibinfo
  {author} {\bibfnamefont {B.~R.}\ \bibnamefont {Safdi}},\ }\bibfield  {title}
  {\bibinfo {title} {Upper {Limit} on the {QCD} {Axion} {Mass} from {Isolated}
  {Neutron} {Star} {Cooling}},\ }\href
  {https://doi.org/10.1103/PhysRevLett.128.091102} {\bibfield  {journal}
  {\bibinfo  {journal} {Phys. Rev. Lett.}\ }\textbf {\bibinfo {volume} {128}},\
  \bibinfo {pages} {091102} (\bibinfo {year} {2022})}\BibitemShut {NoStop}%
\bibitem [{\citenamefont {Yan}\ \emph {et~al.}(2019)\citenamefont {Yan},
  \citenamefont {Sun}, \citenamefont {Peng}, \citenamefont {Guo}, \citenamefont
  {Liu} \emph {et~al.}}]{Yan:2019dar}%
  \BibitemOpen
  \bibfield  {author} {\bibinfo {author} {\bibfnamefont {H.}~\bibnamefont
  {Yan}}, \bibinfo {author} {\bibfnamefont {G.~A.}\ \bibnamefont {Sun}},
  \bibinfo {author} {\bibfnamefont {S.~M.}\ \bibnamefont {Peng}}, \bibinfo
  {author} {\bibfnamefont {H.}~\bibnamefont {Guo}}, \bibinfo {author}
  {\bibfnamefont {B.~Q.}\ \bibnamefont {Liu}}, \emph {et~al.},\ }\bibfield
  {title} {\bibinfo {title} {{Constraining exotic spin dependent interactions
  of muons and electrons}},\ }\href
  {https://doi.org/10.1140/epjc/s10052-019-7442-8} {\bibfield  {journal}
  {\bibinfo  {journal} {Eur. Phys. J. C}\ }\textbf {\bibinfo {volume} {79}},\
  \bibinfo {pages} {971} (\bibinfo {year} {2019})}\BibitemShut {NoStop}%
\bibitem [{\citenamefont {Terrano}\ \emph {et~al.}(2015)\citenamefont
  {Terrano}, \citenamefont {Adelberger}, \citenamefont {Lee},\ and\
  \citenamefont {Heckel}}]{Terrano_Probe_2015}%
  \BibitemOpen
  \bibfield  {author} {\bibinfo {author} {\bibfnamefont {W.~A.}\ \bibnamefont
  {Terrano}}, \bibinfo {author} {\bibfnamefont {E.~G.}\ \bibnamefont
  {Adelberger}}, \bibinfo {author} {\bibfnamefont {J.~G.}\ \bibnamefont
  {Lee}},\ and\ \bibinfo {author} {\bibfnamefont {B.~R.}\ \bibnamefont
  {Heckel}},\ }\bibfield  {title} {\bibinfo {title} {Short-range,
  spin-dependent interactions of electrons: A probe for exotic pseudo-goldstone
  bosons},\ }\href {https://doi.org/10.1103/PhysRevLett.115.201801} {\bibfield
  {journal} {\bibinfo  {journal} {Phys. Rev. Lett.}\ }\textbf {\bibinfo
  {volume} {115}},\ \bibinfo {pages} {201801} (\bibinfo {year}
  {2015})}\BibitemShut {NoStop}%
\bibitem [{\citenamefont {Terrano}\ \emph {et~al.}(2019)\citenamefont
  {Terrano}, \citenamefont {Adelberger}, \citenamefont {Hagedorn},\ and\
  \citenamefont {Heckel}}]{Terrano_constraints_2019}%
  \BibitemOpen
  \bibfield  {author} {\bibinfo {author} {\bibfnamefont {W.~A.}\ \bibnamefont
  {Terrano}}, \bibinfo {author} {\bibfnamefont {E.~G.}\ \bibnamefont
  {Adelberger}}, \bibinfo {author} {\bibfnamefont {C.~A.}\ \bibnamefont
  {Hagedorn}},\ and\ \bibinfo {author} {\bibfnamefont {B.~R.}\ \bibnamefont
  {Heckel}},\ }\bibfield  {title} {\bibinfo {title} {Constraints on axionlike
  dark matter with masses down to ${10}^{\ensuremath{-}23}\text{ }\text{
  }\mathrm{eV}/{c}^{2}$},\ }\href
  {https://doi.org/10.1103/PhysRevLett.122.231301} {\bibfield  {journal}
  {\bibinfo  {journal} {Phys. Rev. Lett.}\ }\textbf {\bibinfo {volume} {122}},\
  \bibinfo {pages} {231301} (\bibinfo {year} {2019})}\BibitemShut {NoStop}%
\bibitem [{\citenamefont {Aprile}\ \emph {et~al.}(2022)\citenamefont {Aprile},
  \citenamefont {Abe}, \citenamefont {Agostini}, \citenamefont
  {Ahmed~Maouloud}, \citenamefont {Althueser} \emph {et~al.}}]{SearchXENONnT}%
  \BibitemOpen
  \bibfield  {author} {\bibinfo {author} {\bibfnamefont {E.}~\bibnamefont
  {Aprile}}, \bibinfo {author} {\bibfnamefont {K.}~\bibnamefont {Abe}},
  \bibinfo {author} {\bibfnamefont {F.}~\bibnamefont {Agostini}}, \bibinfo
  {author} {\bibfnamefont {S.}~\bibnamefont {Ahmed~Maouloud}}, \bibinfo
  {author} {\bibfnamefont {L.}~\bibnamefont {Althueser}}, \emph {et~al.}
  (\bibinfo {collaboration} {XENON Collaboration}),\ }\bibfield  {title}
  {\bibinfo {title} {Search for new physics in electronic recoil data from
  xenonnt},\ }\href {http://dx.doi.org/10.1103/physrevlett.129.161805}
  {\bibfield  {journal} {\bibinfo  {journal} {Phys. Rev. Lett.}\ }\textbf
  {\bibinfo {volume} {129}},\ \bibinfo {pages} {161805} (\bibinfo {year}
  {2022})}\BibitemShut {NoStop}%
\bibitem [{\citenamefont {Gondolo}\ and\ \citenamefont
  {Raffelt}(2009)}]{Gondolo_Solar_2022}%
  \BibitemOpen
  \bibfield  {author} {\bibinfo {author} {\bibfnamefont {P.}~\bibnamefont
  {Gondolo}}\ and\ \bibinfo {author} {\bibfnamefont {G.~G.}\ \bibnamefont
  {Raffelt}},\ }\bibfield  {title} {\bibinfo {title} {Solar neutrino limit on
  axions and kev-mass bosons},\ }\href
  {https://doi.org/10.1103/PhysRevD.79.107301} {\bibfield  {journal} {\bibinfo
  {journal} {Phys. Rev. D}\ }\textbf {\bibinfo {volume} {79}},\ \bibinfo
  {pages} {107301} (\bibinfo {year} {2009})}\BibitemShut {NoStop}%
\bibitem [{\citenamefont {Capozzi}\ and\ \citenamefont
  {Raffelt}(2020)}]{Capozzi_Axion_2020}%
  \BibitemOpen
  \bibfield  {author} {\bibinfo {author} {\bibfnamefont {F.}~\bibnamefont
  {Capozzi}}\ and\ \bibinfo {author} {\bibfnamefont {G.}~\bibnamefont
  {Raffelt}},\ }\bibfield  {title} {\bibinfo {title} {Axion and neutrino bounds
  improved with new calibrations of the tip of the red-giant branch using
  geometric distance determinations},\ }\href
  {https://doi.org/10.1103/PhysRevD.102.083007} {\bibfield  {journal} {\bibinfo
   {journal} {Phys. Rev. D}\ }\textbf {\bibinfo {volume} {102}},\ \bibinfo
  {pages} {083007} (\bibinfo {year} {2020})}\BibitemShut {NoStop}%
\end{thebibliography}%
\end{document}